# Fast MRI Reconstruction Using Deep Learning-based Compressed Sensing: A Systematic Review

Mojtaba Safari[1], Zach Eidex[1] Chih-Wei Chang[1], Richard L.J. Qiu[1], Xiaofeng Yang[1,‡]

[1] Department of Radiation Oncology and Winship Cancer Institute, Emory University, Atlanta, GA 30322, United States of America,

[‡] corresponding Author

Author to whom correspondence should be addressed. email: xiaofeng.yang@emory.edu

**running title:** DL-based CS-MRI

**Abstract:**

Magnetic resonance imaging (MRI) has revolutionized medical imaging, providing a non-invasive and highly detailed look into the human body. However, the long acquisition times of MRI present challenges, causing patient discomfort, motion artifacts, and limiting real-time applications. To address these challenges, researchers are exploring various techniques to reduce acquisition time and improve the overall efficiency of MRI. One such technique is compressed sensing (CS), which reduces data acquisition by leveraging image sparsity in transformed spaces. In recent years, deep learning (DL) has been integrated with CS-MRI, leading to a new framework that has seen remarkable growth. DL-based CS-MRI approaches are proving to be highly effective in accelerating MR imaging without compromising image quality. This review comprehensively examines DL-based CS-MRI techniques, focusing on their role in increasing MR imaging speed. We provide a detailed analysis of each category of DL-based CS-MRI including end-to-end, unroll optimization, self-supervised, and federated learning. Our systematic review highlights significant contributions and underscores the exciting potential of DL in CS-MRI. Additionally, our systematic review efficiently summarizes key results and trends in DL-based CS-MRI including quantitative metrics, the dataset used, acceleration factors, and the progress of and research interest in DL techniques over time. Finally, we discuss potential future directions and the importance of DL-based CS-MRI in the advancement of medical imaging. To facilitate further research in this area, we provide a GitHub repository that includes up-to-date DL-based CS-MRI publications and publicly available datasets - https://github.com/mosaf/Awesome-DL-based-CS-MRI.

**Abbreviations:**

**bSSFP** Balanced steady-state free precession

**CS** compressed sensing

**DC** Data consistency

**DL** Deep learning

**DL-based CS-MRI** Deep learning-based compressed sensing-magnetic resonance imaging

**DTI** Diffusion-tensor imaging

**DWI** Diffusion-weighted imaging

**E2E** End-to-end

**FID** Fréchet inception distance

**FL** Federated learning

**FLAIR** Fluid-attenuated inversion recovery

**GRE** Gradient echo

**MAE** Mean absolute error

**MRA** Magnetic resonance angiography

**MRI** Magnetic resonance imaging

**MSE** Mean square error

**NMSE** Normalized mean square error

**PD** Proton density

**PDFS** Proton density fat-suppression

**PSNR** peak signal-to-noise ratio

**R** Acceleration rate

**RMSE** Root mean square error

**RSS** Root sum of squares

**T1c** T1-w post-contrast MRI

**T1-MPRAGE** T1w magnetization-prepared rapid gradient-echo

**SSIM** Structural similarity index

**SWI** Susceptibility-weighted imaging

# 1 Introduction

Magnetic Resonance Imaging (MRI) is a highly effective medical tool that produces high-quality images of soft tissues in the body. It is widely used in the lesion prognosis and diagnosis, radiation treatment planning, and follow-up examination. However, the Lancet Oncology Commission recently highlighted a severe shortage of MRI and other medical imaging technologies in low-income and middle-income countries (Hricak et al., 2021). This shortage has resulted in 2.5 million deaths worldwide. The installation of MRI scanners remains low globally, with only 7 MRI scanners per million people installed as of 2020 (Y. Liu et al., 2021). This is primarily due to the high cost of installation, operation, and maintenance. Besides, the daily throughput is limited by the long acquisition time required for each MRI scan. The long wait time reduces the number of patients that can be seen in a given day (Murali et al., 2023). Additionally, it increases the likelihood of voluntary and involuntary patient movements (Safari et al., 2023b), causing motion artifacts that affect the accuracy of the images produced. The estimated cost of motion artifacts induced by patient movements is around $364,000 per scanner annually (Slipsager et al., 2020).

MRI requires densely sampled k-space to avoid violating the Nyquist criteria, which results in longer acquisition times for high-resolution images. To reduce the imaging time, $k$-space can be undersampled in the phase encoding direction by increasing the spacing between $k$-space lines and, therefore, covering the field of view in a shorter amount of time, as illustrated in Figure 1 for a Cartesian trajectory. Compressed sensing (CS), also known as compressive sensing or compressive sampling, is a method that aims to reconstruct fully-sampled k-space from undersampled k-space by exploiting the images' sparse representation in a transform domain such as Cosine and Wavelet (Haldar et al., 2010). The CS algorithms optimize a cost function (1) to iteratively reconstruct the image. However, CS algorithms are unable to completely reconstruct the high-frequency texture content of images (Ravishankar and Bresler, 2010), limiting them to acceleration factors between 2.5 and 3 (Guo et al., 2021a). In addition, these iterative techniques inevitably increase reconstruction time.

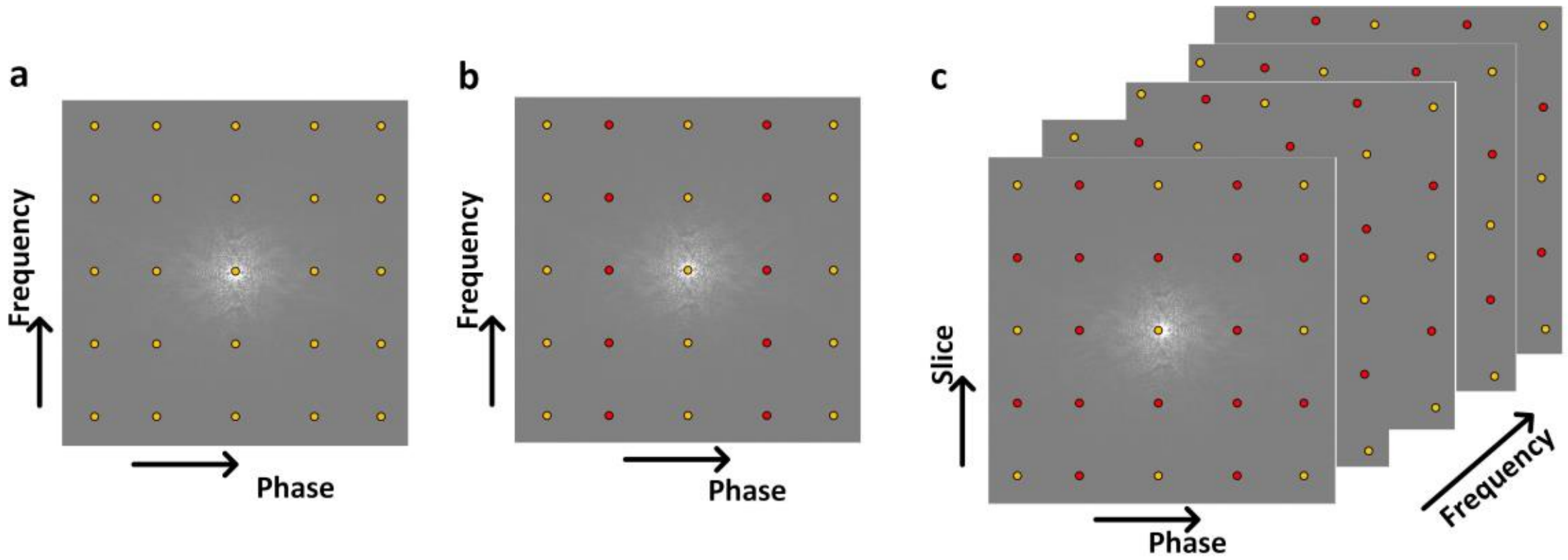

Figure 1: The schematic diagrams of the k-space sampling pattern with a Cartesian trajectory are illustrated for (a) fully sampled data and (b) undersampled 2D MRI data in phase encoding direction with acceleration rate (R) 2 as well as (c) undersampled 3D MRI in both slice and phase encoding direction with $R = 2$. The yellow and red circles indicate sampled and skipped data during the data acquisition.

$$\begin{aligned} \min_{x} \quad &= \|\mathbf{\Psi x}\|_2 \\ s.t. \quad &\|\mathbf{y}_\Omega - \mathbf{E}_\Omega \mathbf{x}\|_2^2 < \epsilon \end{aligned} , \qquad (1)$$

where $\mathbf{E}_\Omega \in \mathbb{C}^{M\times N} \rightarrow \mathbb{C}^M$ given $N > M$ is the encoding operator. It is composed of a coil sensitivity map, a Fourier transform, and a sampling map with the specified pattern $\Omega$. $\mathbf{y}_\Omega \in \mathbb{C}^M$, $\mathbf{x} \in \mathbb{C}^N$, and $\epsilon$ are the undersampled $k$-space measurement, fully sampled image, and the threshold controlling the reconstruction fidelity, which is roughly equal to the expected noise level, respectively.

Parallel imaging is another approach to reduce image acquisition time. Multiple receiver coils are placed in different positions within the scanner to independently collect a portion of the $k$-space data(Deshmane et al., 2012; Larkman and Nunes, 2007). Each coil is most sensitive to the area closest to it (shown in Figure 2a) with the sensitivity relationship encoded in the form of sensitivity maps (shown in Figure 2b). The final image is generated by combining the individual images obtained from different coils by taking their root sum of squares (RSS) and weighting them by the corresponding sensitivity maps.

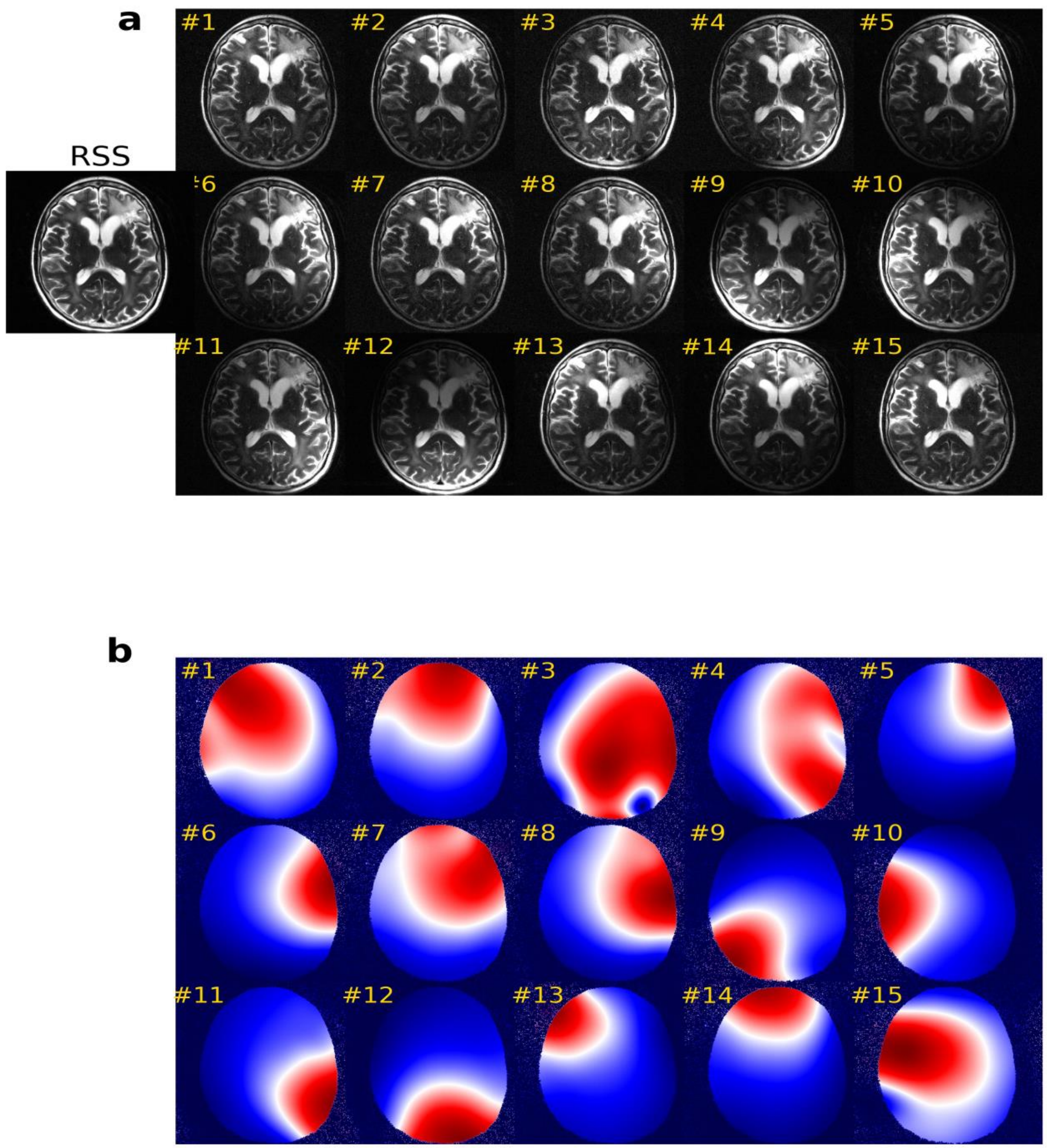

Figure 2: (a) Images and (b) sensitivity maps, estimated by the ESPIRiT approach *(Uecker et al., 2014)*, are illustrated for the first 15 receiver coils. The root sum of squares (RSS) indicates combined weighted receiver coil images by the corresponding sensitivity maps.

Deep learning (DL) algorithms have garnered significant interest in CS-MRI applications. These DL models have demonstrated superior reconstruction performance at a higher acceleration rate compared to traditional non-DL-based CS models (Mardani et al., 2018; Qin et al., 2018). In light of the rapid advancements in this field, we conduct a systematic review, encapsulating the latest developments in state-of-the-art DL-based CS-MRI models. Table 1 provides a concise summary of the most relevant survey articles, revealing technical differences compared to our review. We also note that significant progress has been made in DL-based CS-MRI over the last 2 years, especially with growing interest in diffusion models and the DC layer, so our review better captures current research directions.

Table 1. Related review papers from the DL-based CS-MRI.

| References | Year | Contributions and technical differences |
|---|---|---|
| (Yoon et al., 2023) | 2023 | This review focuses on accelerated musculoskeletal MRI. It does not discuss the statistics on the quantitative metrics and acceleration rates. |
| (Y. Chen et al., 2022) | 2022 | This review details DL algorithms and provided statistics on quantitative metrics. However, our comprehensive review, in addition to those statistics, provided detailed explanations about MRI imaging, such as k-space trajectories, the implications of different sampling patterns, and parallel imaging. Our comprehensive review also discussed the clinical applications of DL-based CS-MRI and provides insights into future directions. |
| (Bustin et al., 2020) | 2020 | The primary consideration is on DL-based CS-MRI for cardiac imaging. In addition to being wider in scope, our review discusses clinical applications of interest to imaging centers, provides relevant statistics, categorizes the DL method used, and lists related references. |
| (Xie and Li, 2022) | 2022 | A review about CS for medical applications. We focus on CS for MRI and categorize based on the study's training method. In addition, our comprehensive review provides details about MRI acquisition and acceleration methods. |

The PubMed database was meticulously searched on February 1st, 2024, using the terms "deep learning reconstruction", "fastMRI", "unrolled optimization", "MRI reconstruction", and "MRI acceleration" for articles published from February 2024 to January 2016. Relevant studies were carefully screened by title and abstract content. Of the 873 publications identified by PubMed, 94 articles were included. Figure 3 illustrates the entire literature screening and selection process.

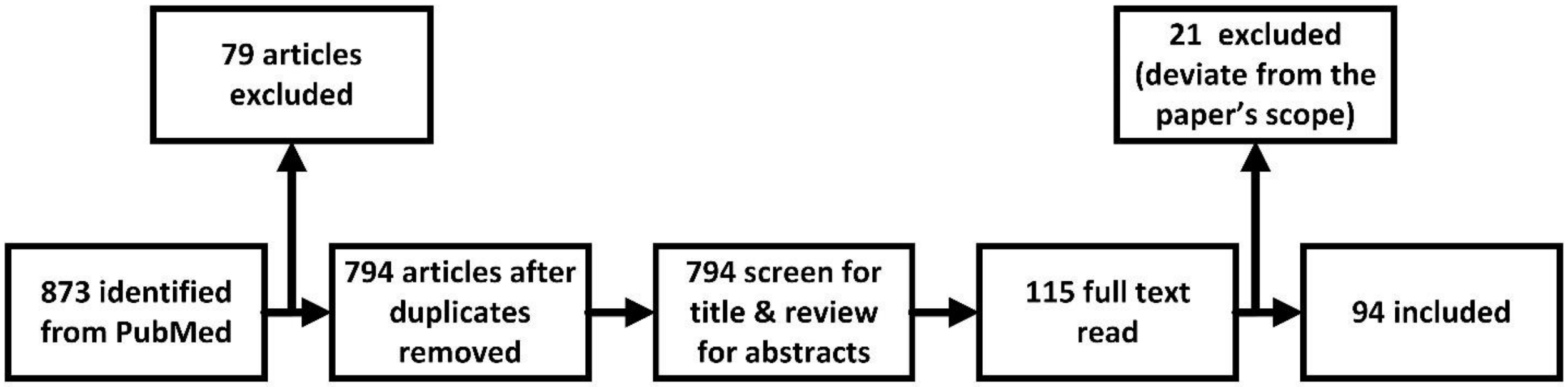

Figure 3: Flowchart of our study selection process.

## 1.1 k-space trajectories

Various trajectories have been developed in MRI for traversing k-space, including Cartesian, spiral, radial, and random trajectories. The Cartesian trajectory, as depicted in Figure 4a, consists of parallel lines equidistant from each other, with each line representing a frequency-encoding readout. The image can be reconstructed using a fast Fourier transform, but each line requires a separate RF pulse, prolonging imaging time.

The radial trajectory, first used by (Lauterbur, 1973) and shown in Figure 4b, consists of spokes radiating from the center, with an oversampling center in k-space that makes it robust to motion artifacts (Maclaren et al., 2013). However, undersampling in the azimuthal direction increases streak artifacts (Xue et al., 2012).

The spiral trajectory shown in Figure 4c was introduced to decrease the MRI acquisition time. It starts at the center of the k-space and spirals outward, similar to radial sampling, and is robust to motion artifacts. However, hardware limitations restrict imaging efficiency and increase image blurring.

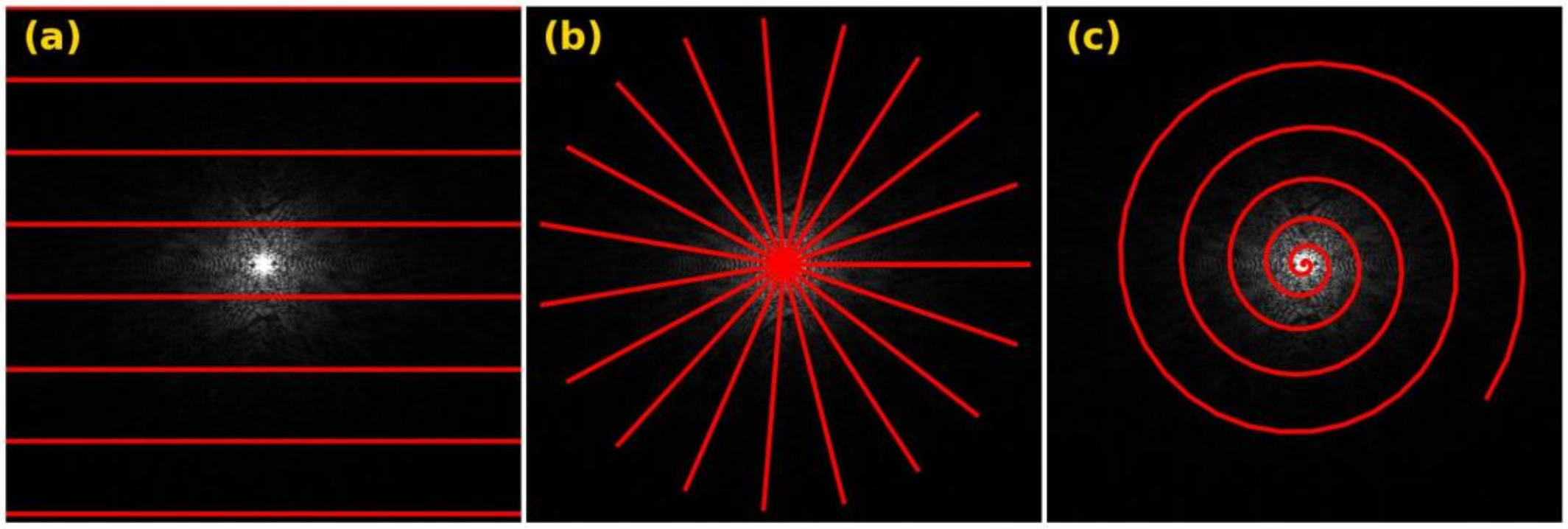

Figure 4: An exemplary k-space with overlaid (a) Cartesian, (b) radial, and (c) spiral trajectories illustrated.

## 1.2 Sampling patterns

To speed up the process of capturing images, a pattern or mask, denoted as $\boldsymbol{\Omega}$ in (1), is utilized to sample k-space. Numerous sampling patterns have been introduced for data acquisition techniques including Cartesian (Safari et al., 2024), Poisson (Slavkova et al., 2023), Gaussian (C. Hu et al., 2021), radial (Terpstra et al., 2023), and spiral patterns. Figure 5 illustrates examples of the different sampling patterns. The Cartesian pattern is typically used for brain and knee data with a Cartesian k-space trajectory while we found the Poisson and Gaussian patterns to be primarily used to train self-supervised models as described in Section 3.4. On the other hand, radial and spiral patterns are mainly used for capturing myocardial and dynamic images that are more likely to have motion artifacts.

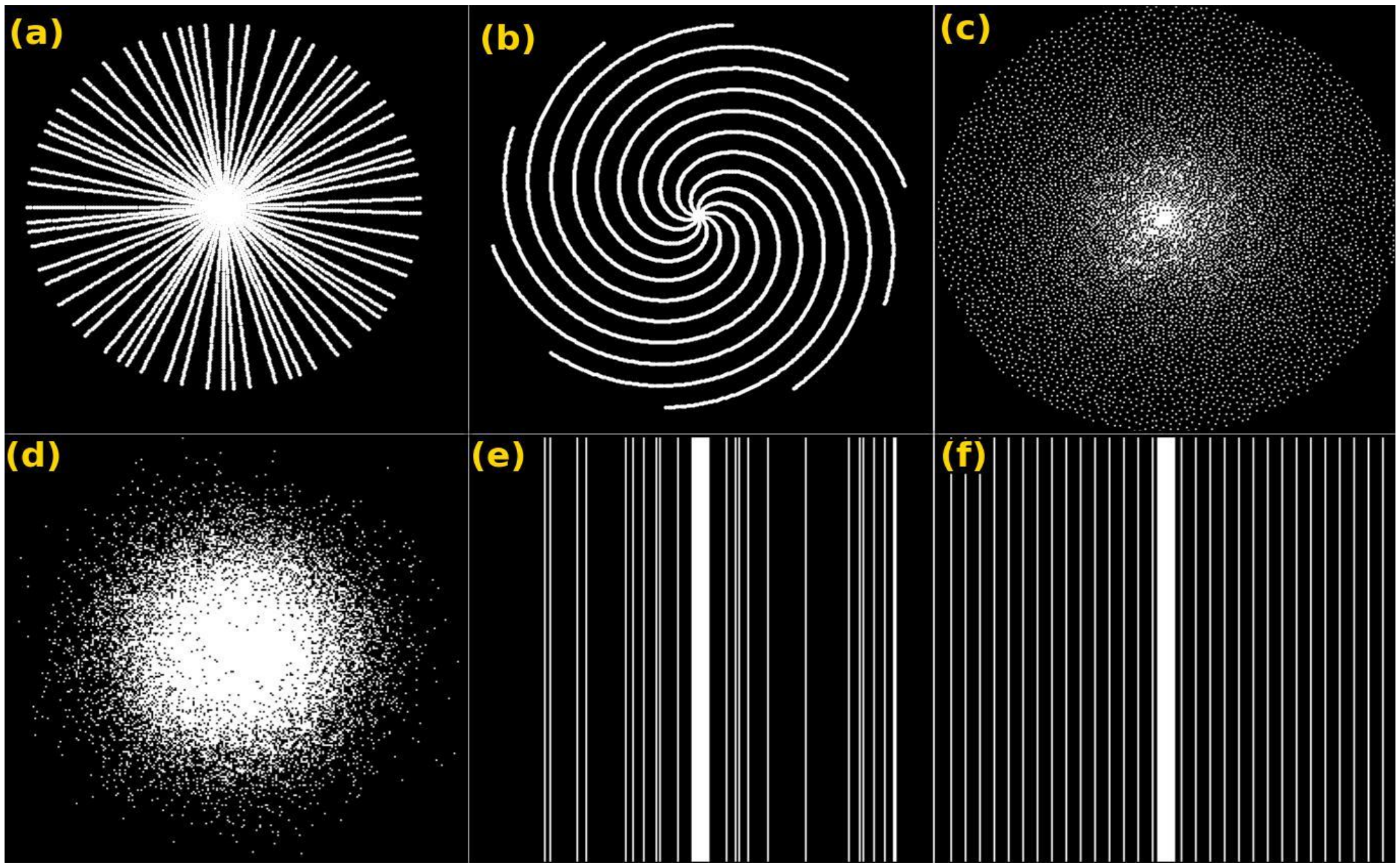

Figure 5: Widely used sampling pattern for k-space undersampling - (a) radial, (b) spiral, (c) Poisson disk, (d) Gaussian, (e) random Cartesian, and (f) equidistant Cartesian.

# 2 Deep learning

## 2.1 Convolutional neural network

Convolutional neural networks, also known as ConvNets, are a type of deep neural networks that are designed to analyze grid-like data such as images and speech (LeCun et al., 1995). They have gained widespread recognition after the success of AlexNet (Krizhevsky et al., 2012) and have since been used to achieve state-of-the-art performance in various medical image processing and analysis tasks. ConvNets typically consist of multiple layers, with each layer including a convolutional operator, batch normalization layers, nonlinear activation functions, and dropout layers. Nonlinear activation functions are used to facilitate the learning of complex functions. Finally, weight regularization and dropout layers are employed to mitigate overfitting. During convolution, trainable convolution kernels slide over the images to extract multiple feature maps called channels.

The network's optimum parameters are computed using a backpropagation algorithm that calculates the gradient of the cost function with respect to the parameters in each layer. Batch normalization layers are crucial in training deep ConvNets to prevent vanishing and exploding gradients. In addition, residual blocks (He et al., 2016) are a popular choice for building advanced ConvNets due to their ability to prevent gradient vanishing and facilitate smoother error surfaces (Li et al., 2018). By incorporating skip-layer connections between input and output, residual blocks can help reduce the risk of local minima. Furthermore, when combined with (batch) normalization layers, residual blocks can effectively address the problems of vanishing and exploding gradients.

## 2.2 Network architectures

### 2.2.1 U-Net

Several deep learning models with different architectures have been proposed to enhance the performance and generalization of ConvNets. Among them, U-Net, with its elegant design that utilizes skip connections between the encoder and decoder, is the best known architecture in computer vision (Ronneberger et al., 2015). It has been extensively exploited in different medical applications such as image synthesis (Han, 2017), segmentation (Dong et al., 2019), and registration (Balakrishnan et al., 2019). In recent years, U-net architectures have incorporated residual and attention layers as a backbone to increase the network's depth and improve performance.

### 2.2.2 Transformer

While ConvNets have been impressive in their results, they are limited by the local context of convolutional operations. To address this challenge, Transformers have emerged as a solution to capture global context (Dosovitskiy et al., 2020), often outperforming ConvNets. However, transformer models are fundamentally very complex and require many trainable parameters and large databases for training which can be a challenge in medical imaging. To mitigate these issue, various variations have been proposed, such as Swin Transformers (Z. Liu et al., 2021), Vision CNN-Transformer (Fang et al., 2022), and ReconFormer (P. Guo et al., 2023), which aim to reduce model size while improving or maintaining performance.

### 2.2.3 Generative adversarial network

Generative adversarial networks (GANs) are implicit methods. Thus, they do implicitly attempt to minimize likelihood function nor attempt to learn latent representation. The GAN, initially introduced in 2014, consists of two networks, generative and discriminator (Goodfellow et al., 2020). The former is trained to generate artificial data samples to approximate the target data distribution, and the latter is simultaneously trained to distinguish the artificial data from real ground truth data. Thus, the discriminators encourage the generator to generate data samples with a distribution similar to the target distribution. Variations of GANs have been developed to perform tasks including image-to-image translation, such as conditional GAN (Mirza and Osindero, 2014), StyleGAN (Karras et al., 2019), CycleGAN (Zhu et al., 2017), and Pix2Pix (Isola et al., 2017). GANs are widely used in medical imaging for tasks such as image registration, image synthesis, and MRI image reconstruction (Quan et al., 2018; Shaul et al., 2020; Yang et al., 2017).

### 2.2.4 Diffusion model

The stable diffusion model, inspired by nonequilibrium thermodynamics, aims to simplify complex and difficult-to-calculate distributions using tractable ones like normal Gaussian distributions (Sohl-Dickstein et al., 2015). This model is comprised of two steps - the forward and reverse processes (Figure 6). During the forward process, Gaussian noise is added to the initial image $x_0$ over $T$ steps until the data at step $T$ becomes normal Gaussian noise $x_T = \mathcal{N}(\mathbf{0}, \mathbf{I})$. In the reverse process, the model learns to recover the original image $x_0$ from its noisy version given at a step $t \in (0, T]$ (Chan, 2024).

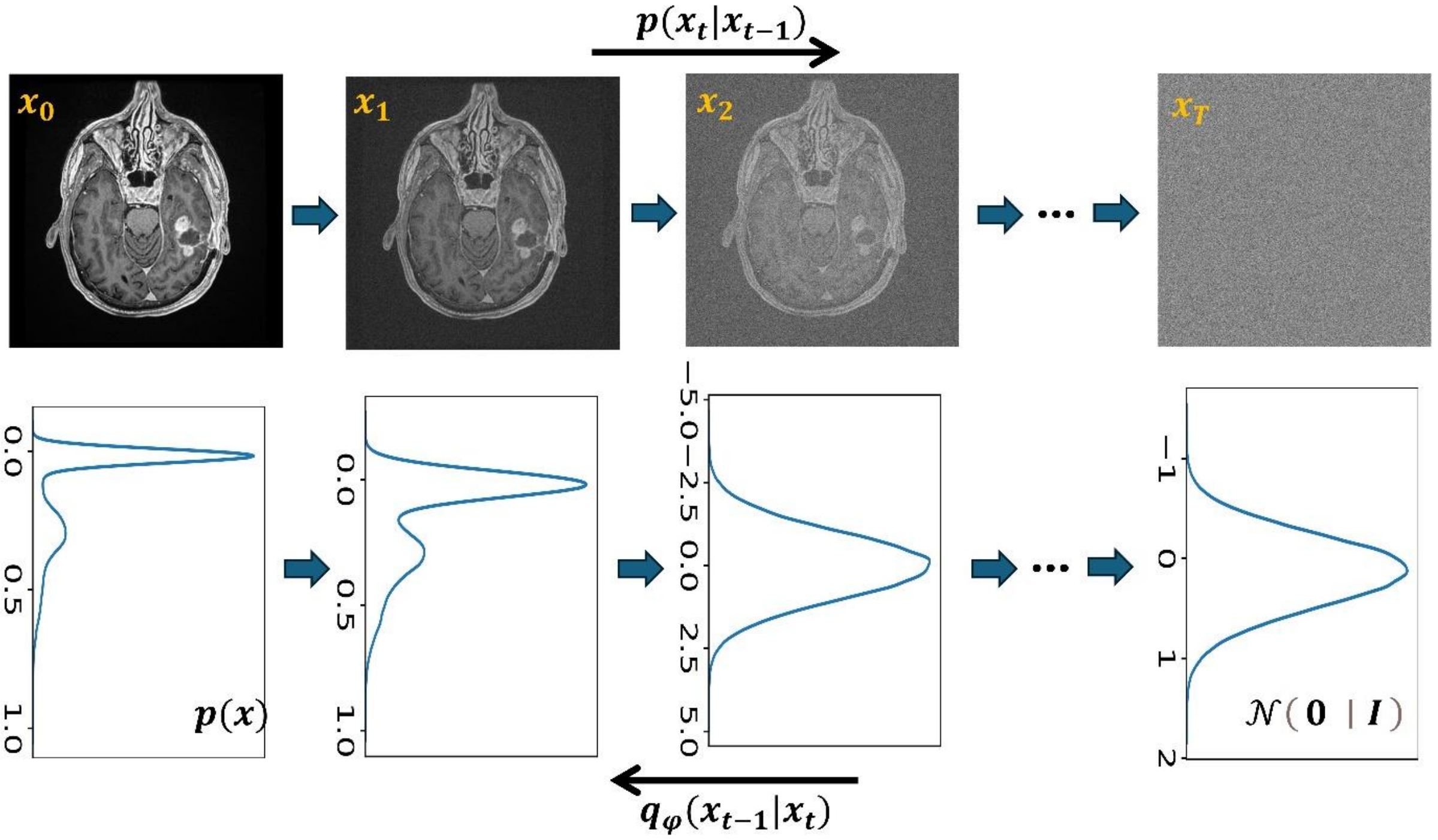

Figure 6: Forward and reverse diffusion processes. The first row indicates data in image space, and the second row indicates the corresponding data distribution. The forward diffusion process adds Gaussian noise in $T$ steps in a controlled way to produce normal Gaussian noise in step $T$. The reverse diffusion process requires a model parametrized by $\varphi$ to learn input $x_0$ from noise-corrupted image $x_t$ in a given step $t$.

Stable diffusion models have been employed for medical imaging tasks such as denoising (Pan et al., 2023b), synthesis (Pan et al., 2023a), MRI distortion reduction (Safari et al., 2023b), and MRI image reconstruction (Chung and Ye, 2022; Güngör et al., 2023).

# 3 Deep learning for MRI reconstruction

The framework for DL-based CS-MRI can be divided into two main categories: data-driven and physics-driven models. Within these categories, there are two types of models: end-to-end and unroll. End-to-end models take in zero-filled k-space and output fully-sampled k-space. They typically utilize a regularization term listed in Table 2 to enforce the uniqueness of the reconstructed images. On the other hand, unroll models are more complex and further classified into two types: unroll optimization and closed-form models known for the "data consistency (DC) layer." Unroll optimization models iteratively optimize the reconstruction process, while DC layer models use a closed-form equation to ensure data consistency. These models are utilized in various training scenarios, including federated learning and self-supervised training. For a comprehensive understanding of DL-based CS-MRI methods and their corresponding components and features, please refer to Figure 7.

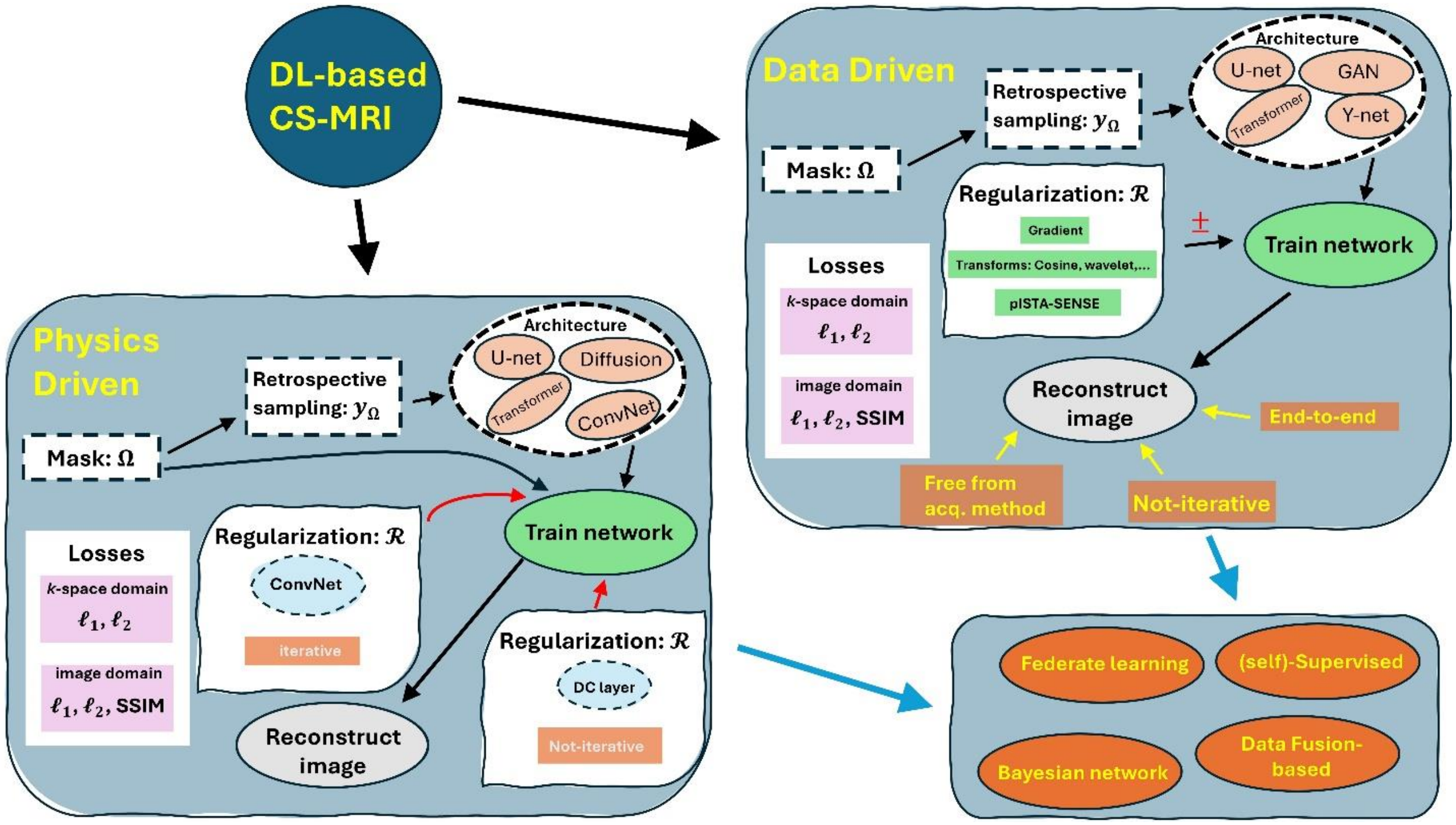

Figure 7: An overview of seven categories of DL-based CS-MRI methods.

Figure 8 shows a stacked chart of the number of publications since 2018 by category. The total number of publications has grown exponentially in recent years and interest in the DC layer method continues to increase while interest in unrolling optimization and end-to-end approaches remains consistent over time.

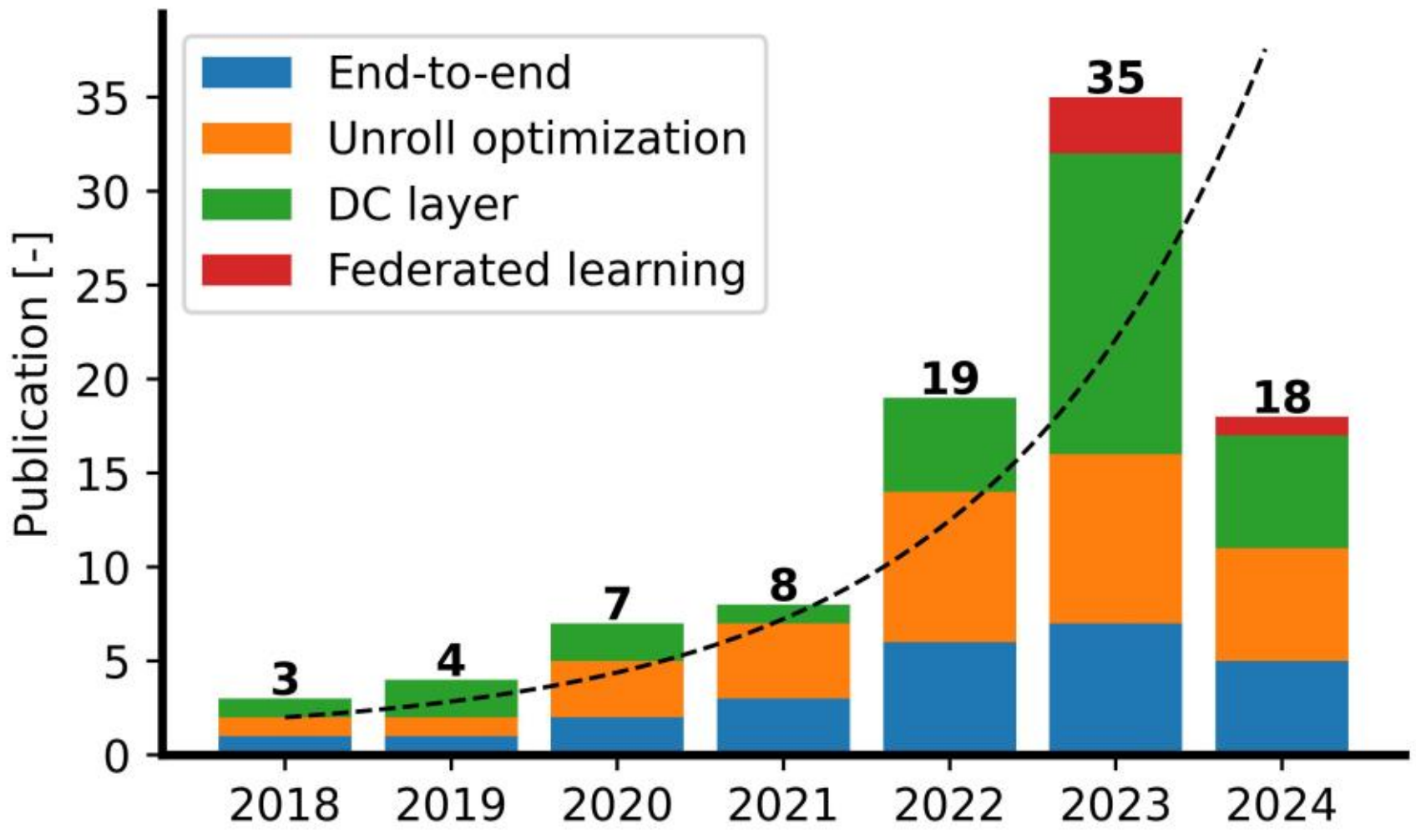

Figure 8: Overview of publications in DL-based CS-MRI over time. The dashed line indicates the general trend plotted using $1 + \exp(0.62 \times t)$ where $t$ is defined in years.

## 3.1 End-to-end models

DL end-to-end models are specifically designed to tackle the CS-MRI problem without enforcing any data acquisition model. To achieve this, these models rely on a neural network to accurately predict fully sampled data from undersampled data (B. Levac et al., 2023; Mardani et al., 2018; Yang et al., 2016). Additionally, these models are trained using various regularization techniques that help address the ill-posed inverse problem. Table 2 lists common regularization techniques utilized in these models.

Table 2: The widely used regularization terms in deep CS-MRI are summarized.

| Traditional | DL-based model | Regularizer | Reference |
|---|---|---|---|
| Dictionary learning | ADMS | $\|\alpha\|_1$ | (Cao et al., 2020) |
| Field of expert | VN | $\sum_i \langle \Phi_i(K_i x), 1 \rangle$ | (Hammernik et al., 2018) |
| pISTA-SENSE | pISTA-SENSE-ResNet | $\|\Psi x\|_1$ | (Lu et al., 2020) |
| Total variation | -, RELAX | $\|\nabla x\|_1$ | (F. Liu et al., 2021; Sun et al., 2017) |
| Sparse and low-rank model | ODLS | $\|\mathcal{H}x\|_*$ | (Z. Wang et al., 2022) |

**Dictionary learning** learns a latent representation $\alpha$ of the input image $x$ where the $\ell_1$ norm enforces it to be sparse.

**Field of expert** is composed of the convolution kernel $K_i$ and $\Phi_i$, which they are learned from data.

**pISTA-SENSE** is a projected iterative soft-thresholding algorithm that solves (1) iteratively where the transform $\Psi$ enforces the sparsity of the reconstructed image $x$.

**Total variation** enforces image smoothness by minimizing the image gradient variations.

**Sparse and low-rank model** minimize the nuclear norm $\|\mathcal{H}x\|_*$ where $\mathcal{H}$ is the Hankel matrix.

The end-to-end approach employs the same baseline models that are used for image-to-image translation, such as the U-nets (Hyun et al., 2018; Xiang et al., 2018), Swin transformers (Huang et al., 2022), and GANs (Shitrit and Riklin Raviv, 2017; Zhao et al., 2023). However, they require a larger sample dataset than unrolling CS-MRI models and tend to predict images with synthetic data.. Table 3 provides a list of selected references that used end-to-end DL models to solve DL-based CS-MRI algorithms.

Table 3: Overview of supervised end-to-end models to predict the fully sampled images.

| DL Model | Dataset | Region | Modality | sample size | R | Ref. | Code[1] |
|---|---|---|---|---|---|---|---|
| GAN | fastMRI[2] (Zbontar et al., 2018) | Knee | PDFS,PD | 299,300 | 4,6 | (Narnhofer et al., 2019) | No |
| GAN | IXI[3]<br>NAMIC[4] | Brain | T2,PD<br>T1,T2 | 72<br>20 | 2,3,4 | (Lyu et al., 2020) | No |
| ResNet | (Hammernik et al., 2018) | knee | PD | 15 | 5,7,9 | (Lu et al., 2020) | Yes |
| U-net | Institutional[5] | Knee | T1$\rho$-map, T2-map | 10 | 2,4,5,6 | (Li et al., 2023) | No |
| GAN | MICCAI 2013[6] | Brain | T1 | 150 | 2,3.3,10 | (Gao et al., 2023) | No |
| | ASC[7] | Cardiac | T1c | 100 | | | |
| | fastMRI | Knee | PD | 96 | | | |
| Y-net | Own | Brain | T1,T2 | 37 | 2,3,4,5,6,8 | (Do et al., 2020) | No |
| U-net | Own | Knee | T1 | 10 | 3,4,5 | (Ayde et al., 2022) | No |
| SwinGAN | IXI<br>MICCAI 2013<br>MRNet | Brain<br>Brain<br>Knee | T1<br>NS[8]<br>NS | 573<br>2480[9]<br>2480[9] | 20,30,50 | (Zhao et al., 2023) | Yes |
| Dense-Unet | MICCAI 2016 | Brain | T2 | 5 | 2,4,8 | (Xiang et al., 2018) | No |
| U-net | Institutional | Cardiac | bSSFP,GRE | 80 | NS | (J. Wang et al., 2024) | No |
| GAN (Pix2pix) | HCP<br>Institutional | Brain | T1,T2<br>T2-map | 20<br>NS | 8 | (Meng et al., 2021) | No |
| GAN | Institutional | Cardiac[10] | DTI | 30 | 2.6 | (Liu et al., 2023) | No |
| DeepADC-Net | Institutional | Animal study | DWI | 183 | 4,8 | (Li et al., 2024) | Yes |
| NPB-REC | fastMRI | Brain, Knee | NS | 5847,1167 | 4,8,12 | (Khawaled and Freiman, 2024) | Yes |
| SCU-Net | Institutional | Brain | T1,T2,T2FLAIR | 180 | 1.78,2.39, 2.91, 3.33,3.64, 4, 4.27,4.44, 4.71 | (Jin and Xiang, 2023) | No |
| DuDReTL U-net | Uk Biobank[11] | Cardiac | NS | 9032 | 20 | (Hong et al., 2023) | Yes |
| iRAKI | fastMRI<br>Institutional | Brain | T1,T2,T1c,T2FLAIR<br>T1,T2,T1FS | 10<br>3 | 4,5 | (Dawood et al., 2023) | Yes |
| MA-RECON | fastMRI<br>MRIdata[12] | Brain<br>Knee | T2FLAIR<br>PD | 1000[9]<br>41872[9] | 4,8 | (Avidan and Freiman, 2024) | Yes |
| Diffusion model | fastMRI | Knee | PD | 973 | 10,12 | (Cao et al., 2024) | Yes |

[1] Whether the original implementation is publicly available,

[2] https://fastmri.med.nyu.edu/,

[3] https://brain-development.org/ixi-dataset/,

[4] https://insight-journal.org/,

[5] The study used its Institutional dataset,

[6] MICCAI 2013 grand challenge dataset: https://wiki.cancerimagingarchive.net/display/Public/NCI-MICCAI+2013+Grand+Challenges+in+Image+Segmentation,

[7] Atrial Segmentation Challenge: https://www.cardiacatlas.org/atriaseg2018-challenge/,

[8] Information not available from the publication,

[9] Number of image slices was specified,

[10] *Ex-vivo* study

[11] https://www.ukbiobank.ac.uk/

[12] https://www.mridata.org/

## 3.2 Unroll model

### 3.2.1 Unroll optimization

Unroll CS-MRI models combine DL with a data acquisition model to solve an optimization problem iteratively. This optimization is given by Equation 2:

$$\hat{x} = argmin \underbrace{\|y_\Omega - E_\Omega x\|_2^2}_{Data\ consistency} + \lambda \underbrace{\mathcal{R}(x)}_{Regularization} \qquad (2)$$

In this equation, $\lambda > 0$ is a scalar regularization weight that balances between the data consistency and regularization terms. The data consistency term ensures that the reconstructed images $\hat{x}$ are similar to the given undersampled $y_\Omega$, thus enforcing data fidelity. The regularization term helps solve the ill-posed CS-MRI problem by imposing sparsity on the solution to guarantee the uniqueness of the reconstructed images, $\hat{\mathbf{x}}$ (Donoho, 2006).

DL models, particularly ConvNets, are used heavily to learn the regularization term through an unroll training scheme. Similarly, the prior DL-based regularizes encode prior knowledge about the reconstructed images, such as sparsity. The unroll CS-MRI DL models have shown to outperform end-to-end methods using a network with a smaller number of trainable parameters (Aggarwal et al., 2018; Geng et al., 2023; Liu et al., 2022; Qiao et al., 2023; Qu et al., 2024).

However, the iterative nature of the unroll optimization method may increase the computation time during both training and inference steps as both the network's weights and data consistency terms are simultaneously updated. Table 4 provides a list of references that used unroll optimization models to solve DL-based CS-MRI algorithms.

Table 4: Overview of supervised unroll optimization to predict the fully sampled images.

| DL Model | Dataset | Region | modality | sample size | R | Reference | Code |
|---|---|---|---|---|---|---|---|
| ConvNet | IXI<br>fastMRI | Brain<br>Knee | T1,T2<br>PD | 50<br>1025 | 2.5, 3.3, 5 | (R. Liu et al., 2020) | Yes |
| ConvNet | Institutional<br>fastMRI | Brain<br>Knee | T1<br>PD, PDFS | 5<br>60 | 4,5,6.7 | (D. Guo et al., 2023) | No |
| RG-Net | Institutional | Brain | T1$\rho$ | 8 | 17 | (Huang et al., 2017) | No |
| ConvNet | Institutional | Brain | SWI | 117 | 5,8 | (Duan et al., 2022) | No |
| ConvNet | HCP (Van Essen et al., 2013) | Brain | T1w,T2w | 1200 | NS[1] | (Zufiria et al., 2022) | No |
| ConvNet | Institutional | Cardiac, Abdominal | NS | 20,16 | 3,4,5 | (Zhou et al., 2019) | No |
| RecurrentVarNet | fastMRI<br>fastMRI<br>Calgary-Campinas (Beauferris et al., 2022) | Brain<br>Knee<br>Brain | T1c, T1, T2,T2-FLAIR<br>PD<br>T1w | 5846<br>1172<br>67 | 2,4,6 | (Yiasemis et al., 2024) | No |
| ConvNet | (Hammernik et al., 2018)<br><br>Institutional | Knee<br><br>Brain | PD<br><br>T1 | 20<br><br>8 | 3,4,6 | (Hammernik et al., 2018) | No |
| UFLoss | fastMRI<br><br>MRIdata | Knee<br><br>Knee | PDFS, PD<br><br>PDFS, PD | 380<br><br>20 | 5<br><br>8 | (K. Wang et al., 2022) | Yes |
| DFSN | SRI24 (Rohlfing et al., 2010)<br><br>MRBrainS13(Mendrik et al., 2015)<br><br>NeoBrainS12 (Išgum et al., 2015) | Brain | PD,T1,T2<br><br>T1,T2-FLAIR<br><br>T1,T2 | 36<br><br>20<br><br>175 | 5, 10 | (Sun et al., 2019) | No |
| ConvNet | Institutional | Brain | T1map | 3 | 8,12,18, 36 | (Slavkova et al., 2023) | Yes |
| JDL | Institutional | Brain | T1,T2,PD,T2 FLAIR | 8 | 4,8 | (Ryu et al., 2021) | Yes |
| pFISTA-DR | Institutional | Brain | T1,PD | 200 | 5,7,10 | (Qu et al., 2024) | No |
| U-net | fastMRI | Knee | PD, PDFS | 20 | 4,6 | (Qiao et al., 2023) | Yes |
| ConvNet | fastMRI<br><br>Institutional | Brain | T1,T2,T2-FLAIR | 120<br><br>3 | 1.8,2.5, 3.5,4 | (Pramanik et al., 2023) | No |
| ConvNet | NAMIC[2]<br><br>MRBrainS[3] | Brain<br><br>Brain | T1,T2<br><br>T2,T2-FLAIR | 5<br><br>7 | 8 | (X. Liu et al., 2021) | No |
| Dictionary learning | Institutional | Cardiac | - | 19 | 2,4,8 | (Kofler et al., 2023) | Yes |
| PD-PCG-Net | fastMRI | Knee | PD, PDFS | 484,489 | 4 | (Kim and Chung, 2022) | No |
| RUN-UP | Institutional | Brain, Breast | DTI, DWI | 14,6 | 100[4] | (Y. Hu et al., 2021) | No |
| ConvNet | Institutional | Cardiac | NS | 22 | 10,12,14 | (Sandino et al., 2021) | No |
| CEST-VN | Institutional | Brain | CEST | 54 | 3,4,5 | (Xu et al., 2024) | Yes |
| Diffusion model | GLIS-RT | Brain | T1c | 230 | 1.25,1.66,2.5,5 | (Safari et al., 2024) | No |

[1] It is not specified in the manuscript

[2] http://hdl.handle.net/1926/1687

[3] https://mrbrains18.isi.uu.nl/

### 3.2.2. Data consistency layer

The more popular approach is to train the unrolled models similarly to the end-to-end models as follows:

$$\hat{x}_{f_\psi} = argmin \left\| x - f_\psi(x_\Omega) \right\|_1 + \lambda \underbrace{\| y_\Omega - E_\Omega x \|_2^2}_{Data\ consistency}, \qquad (3)$$

where $f_\psi$ is a DL model that maps the undersampled input images $x_\Omega$ to reconstruct fully sample images $x$. The DL reconstruction and data consistency operate on the image domain and $k$-space domain, respectively. Although the DL part is trained without incorporating a priori information, the second term discourages the DL first part from updating the $k$-space parts that were not sampled (Schlemper et al., 2017). The closed form for (3) is as follows (Qin et al., 2018):

$$\hat{x} = \begin{cases} \hat{x}_{f_\psi}(k) & if \quad k \notin \Omega \\ \dfrac{\hat{x}_{f_\psi}(k) + \lambda_0 x_\Omega(k)}{\lambda_0 + 1} & Otherwise \end{cases} \quad (4)$$

This closed form is a computational layer called the DC layer at the end of DL models. The DC layer is a crucial part of the CS-MRI DL model, playing an important role in reconstructing images (Cheng et al., 2021; Korkmaz et al., 2023). The DC layer allows for a flexible design of the DL model when it is added to U-net (Murugesan et al., 2021), transformers (Wu et al., 2023), stable diffusion model (Cao et al., 2024), and so on. Table 5 summarizes the DL-based CS-MRI trained under the DC layer framework.

Table 5: Overview of supervised models that used the data consistency to predict the fully sampled images.

| DL Model | Dataset | Region | Modality | sample size | R | Ref. | Code |
|---|---|---|---|---|---|---|---|
| KTMR | Institutional | Brain | T1c, MRA | 17 | 2,2.5,3.3,5,10 | (Wu et al., 2023) | Yes |
| DCT-net | fastMRI<br>Calgary | Brain<br>Knee | T1<br>PD | 973<br>25 | 4,8 | (B. Wang et al., 2024) | No |
| CTFNet | Institutional | Cardiac | bSSFP | 48 | 8,16,24 | (Qin et al., 2021) | Yes |
| EDAEPRec | Institutional | Brain | T2 | 7 | 3.3,4,5,6.7,10 | (Q. Liu et al., 2020) | No |
| GFN | Institutional | Brain | T1MPRAGE,T2FLAIR,TOF | 30,50,80 | 3.3,5,10 | (Dai et al., 2023) | No |
| PC-RNN | fastMRI | Knee<br>Brain | PD, PDFS<br>T1,T2 | 973<br>4469 | 4,6 | (E. Z. Chen et al., 2022) | Yes |
| ResNet Unet | Institutional | Knee | T1 | 360 | 6 | (Wu et al., 2019) | No |
| CNF | fastMRI | Knee, Brain | PD, T2 | 20,8 | 4 | (Wen et al., 2023) | Yes |
| stDLNN | Institutional | Abdomen | GRE | 8 | 4,8,16,25 | (Wang et al., 2023) | Yes |
| MODEST | Institutional | Cardiac | | 28 | 3.7,7.4,14.8 | (Terpstra et al., 2023) | Yes |
| Deepcomplex MRI | Institutional<br>(Hammern | Brain<br>Knee | T1,T2,PD<br>PD | 22<br>20 | 4,5,10 | (Wang et al., 2020) | No |
| | ik et al., 2018) | | | | | | |
| AdaDiff | fastMRI<br>IXI | Brain | T1, T2, PD<br>T1,T2,T2-FLAIR | 66<br>420 | 4,8 | (Güngör et al., 2023) | Yes |
| McSTRA | fastMRI | Knee | PD, PDFS | 584, 588 | 4,6,8,10,12 | (Ekanayake et al., 2024) | No |
| DC-RSN | Cardiac<br>Kirby<br>Calgary<br>MRBrain<br>Hammernick | Cardiac,<br>Brain,<br>Brain,<br>Brain,<br>Knee | Cardiac,<br>T1-MPRAGE<br>T1<br>T2-FLAIR<br>PDF, PD, T1, T2, T2FS | 200<br>42<br>45<br>7<br>25 | 4, 5 | (Murugesan et al., 2021) | No |

## 3.3 Federated learning

Federated learning (FL) is a promising framework that enables the collaborative training of learning-based models across multiple institutions without the need for sharing local private data (Yang et al., 2019). The objective of FL models is to learn a global model by taking the average of local models (McMahan et al., 2017) or by ensuring the proximity of local models to the global model (Li et al., 2020). When applied to MRI image reconstruction, the FL offers unique advantages tailored to the specific challenges and requirements as follows:

- MR images often contain sensitive patient information that needs to be protected. FL enables MRI models to be trained directly on the devices where the images are acquired, without the need to transmit patient data to a centralized location. This decentralization of data ensures privacy and confidentiality of patient information is maintained.
- MRI machines can vary in their hardware specifications and imaging protocols, which can lead to challenges in standardizing image reconstruction algorithms. However, FL accommodates this heterogeneity by allowing models to be trained collaboratively across different types of MRI machines, ensuring that the reconstruction algorithms are robust and adaptable to various configurations.

It is worth noting that FL models are predominantly supervised and have been developed under the end-to-end and unroll model frameworks, which have shown promising results in various applications.

Table 6. Overview of supervised models that were trained with federated learning.

| DL Model | Training framework | Dataset | sample size | Ref. | Code |
|---|---|---|---|---|---|
| FL-MRCM | End-to-End | fastMRI,<br>HPKS(Jiang et al., 2019)<br>IXI,<br>BraTS(Menze et al., 2014) | 3443<br>144<br>434<br>494 | (Guo et al., 2021b) | Yes |
| FedMRI | End-to-End | fastMRI<br>BraTS<br>SMS (Feng et al., 2021)<br>uMR (Feng et al., 2021) | 2134<br>385<br>155<br>50 | (Feng et al., 2022) | Yes |
| ACM-FedMRI | DC guided | fastMRI<br>BraTS<br>IXI<br>Institutional | 3443<br>494<br>526<br>150 | (Lyu et al., 2023) | No |
| FedGIMP | DC guided | fastMRI<br>BraTS<br>IXI<br>Institutional | 51<br>55<br>55<br>10 | (Elmas et al., 2022) | Yes |
| Unrolled FedLrn | Unroll | fastMRI, MRIdata | NS | (B. R. Levac et al., 2023) | Yes |

## 3.4 Self-supervised learning

In contrast to supervised learning methods that necessitate fully sampled ground truth images, self-supervised models alleviate this requirement and are often trained using unrolling optimization techniques. The training framework of self-supervised algorithms does not mandate fully sampled ground truth images. This approach is particularly advantageous in scenarios where obtaining fully sampled data without distortions is challenging, such as myocardial perfusion with the patient's involuntary movements, which cause motion artifacts (Haji-Valizadeh et al., 2018). Self-supervised methods draw samples from the undersampled pattern $\Omega$ provided in (3) to generate a new pattern $\Lambda$ and $\Theta$, where $\Omega = \Lambda \cup \Theta$ and $\Theta = \Omega \setminus \Lambda$. The former is used to train the DL model, while the latter is utilized to compute the loss (Feng et al., 2023; Heydari et al., 2024).

Table 7: Overview of self-supervised models to predict the fully sampled images.

| DL Model | Dataset | Region | modality | sample size | R | Ref. | Code |
|---|---|---|---|---|---|---|---|
| DC-SiamNet | IXI<br>MRINet | Brain<br>Knee | T1, T2, PD<br>PD | 473<br>1250 | 4, 5 | (Yan et al., 2023) | No |
| Noise2Recon | MRIdata<br>fastMRI | Knee<br>Brain | PDFS<br>T2,T2-FLAIR,T1,T1c | 19<br>603 | 12,16 | (Desai et al., 2023) | Yes |
| DDSS | HCP | Brain | T1 | 505 | 2,4 | (Zhou et al., 2022) | No |
| SSDU | Institutional<br>fastMRI<br>fastMRI | Brain<br>Knee<br>Brain | T1MPRAGE<br>PD,PDFS<br>T2 | 19<br>35,25<br>10 | 4,6,8 | (Yaman et al., 2020) | Yes |
| Multi-mask SSDU | Institutional<br>MRIDdata | Brain<br>Knee | T1MPRAGE<br>PD | 9<br>20 | 8,12 | (Yaman et al., 2022) | No |
| RELAX | Institutional | Brain, Knee | T1 and T2 maps | 21 | 5 | (F. Liu et al., 2021) | No |
| Joint MAPLE | Institutional | Brain | T1 and $T2^*$ maps | 2 | 16,25 | (Heydari et al., 2024) | Yes |

## 3.5 Assessment

In the previous sections, we covered three major DL-based CS-MRI training approaches: end-to-end, unroll optimization, and the DC layer. While these methods can be applied to both supervised and self-supervised frameworks, the unroll optimization and DC layer approaches are typically used for self-supervised training. For your convenience, we summarized the advantages and disadvantages of each approach in Table 8.

Table 8: A bullet list of the pros and cons of each training framework is provided.

| DL method | Pros | Cons |
|---|---|---|
| End-to-end | 1. Free from MRI data acquisition method<br>2. Easy to employ models proposed for image synthesis and segmentation | 1. Requires a big dataset<br>2. Likely to add unwanted image structure, specifically for higher acceleration rate<br>3. Limited generalization to out-of-distribution data |
| Unroll optimization | 1. Requires smaller datasets compared with the end-to-end method<br>2. More likely to generalize better<br>3. Performs well using a ConvNet with small number of trainable parameters | 1. Iterative method<br>2. Requires image acquisition knowledge<br>3. Requires undersampling pattern in inference time |
| DC Layer | 1. Requires smaller dataset than the end-to-end method<br>2. Provides closed form equation<br>3. Easy to implement | 1. Requires undersampling pattern in inference time |

# 4 Evaluation metrics

## 4.1 Quantitative metrics

When testing CS-MRI models, the quality of reconstructed images is quantitatively assessed by comparing them with the ground truth. Retrospective undersampling of the k-space allows the reporting of indices to quantify the quality of the reconstructed images.

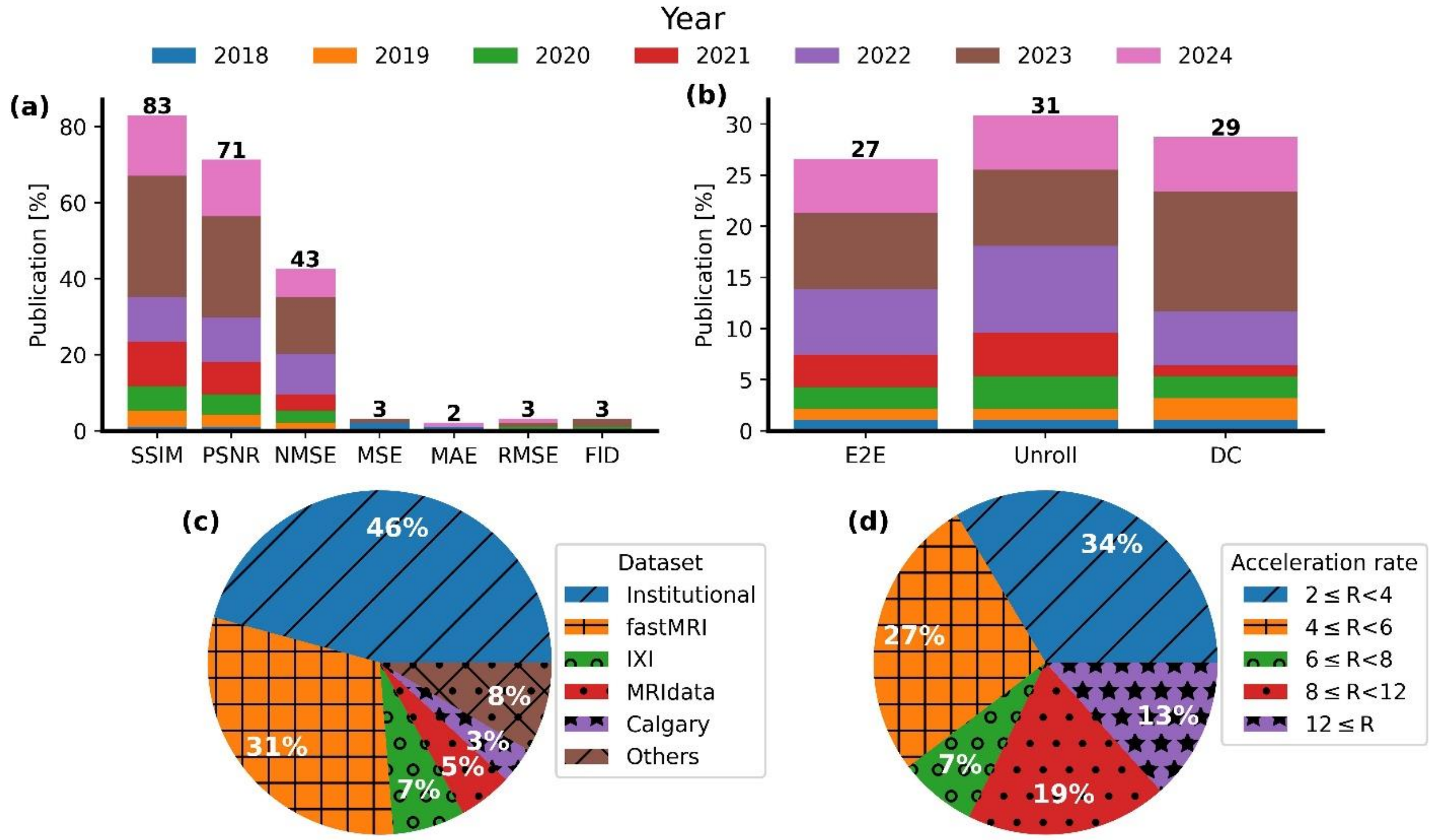

Figure 9: This graph illustrates (a) The metric used in different studies, (b) the training method used in different studies, (c) the dataset used for training used in the studies, and (d) the acceleration rate of the undersampling. Abbreviations: SSIM: structural similarity index; PSNR: peak signal-to-noise ratio; NMSE: normalized mean square error; NRMSE: normalized root mean square error; MSE: mean square error; MAE: mean absolute error; RMSE: root mean square error; and FID: Fréchet inception distance.

Most studies quantitatively compare predicted images with ground truth images. As indicated in Figure 9a, most of these studies use the structural similarity (SSIM) index or peak signal-to-noise ratio (PSNR) to compare reconstructed image $\hat{x}$ with ground truth image $x$ as follows:

$$\mathrm{PSNR}(\hat{x}, x) = \log_{10} \frac{L^2}{\frac{1}{N} \parallel \hat{x} - x \parallel_2^2} \ , \qquad (5)$$

where $\|\cdot\|_2$ is the squared Euclidean distance, $N$ is the number of images' voxels, and $L$ is the maximum voxel intensity of $x$. A higher PSNR indicates a better reconstruction. The logarithmic operator quantifies image quality that closely aligns with human perception (Safari et al., 2023a).

$$\mathrm{SSIM}(\hat{x}, x) = \frac{(2\mu_{\hat{x}}\mu_x + c_1)(2\sigma_{\hat{x}x} + c_2)}{(\mu_{\hat{x}}^2 + \mu_x^2 + c_1)(\sigma_{\hat{x}}^2 + \sigma_x^2 + c_2)}, \qquad (6)$$

where $\mu_{\hat{x}}$ and $\mu_x$ are the average voxel intensities in $\hat{x}$ and $x$, $\sigma_{\hat{x}}$ and $\sigma_x$ are the variance, and $\sigma_{\hat{x}x}$ is the covariance between $\hat{x}$ and $x$. The constants $c_1$ and $c_2$ stabilizes the division, which usually are $c_1 = (k_1L)^2$ and $c_2 = (k_2L)^2$. SSIM ranges between -1 and 1, with the best similarity achieved by an SSIM equal to one.

The normalized mean square error (NMSE) has become more popular since 2022 to quantify the quality of reconstructed images. The NMSE is defined as

$$\mathrm{NMSE}(\hat{\mathrm{x}}, \mathrm{x}) = \frac{\| \hat{\mathrm{x}} - \mathrm{x} \|_2^2}{\| \mathrm{x} \|_2^2}, \qquad (7)$$

Smaller NMSE values indicate better image reconstruction. However, it favors image smoothness rather than sharpness.

However, other metrics such as root mean square error, mean square error, mean absolute error, and Fréchet inception distance are rarely used, especially after a recommendation made in 2018 by Zbontar, Jure, *et al.* (Zbontar et al., 2018) (see Figure 8a).

Regarding the training methods, the Unroll models, including the Unroll optimization and DC layer, are the most commonly used, with a growing use of the DC layer since 2020, which is expected to continue this trend in the future. More details about these trends are illustrated in Figure 9b.

Our systematic review found that around 46% of studies used their own dataset, while fastMRI was the most frequently used public dataset. The majority of private datasets use single-coil images compared to the later's raw multi-coil raw 2D $k$-space data. The fastMRI dataset consists of three imaging regions: the brain, pelvis, and knee regions. Less commonly used datasets include IXI, MRIdata, Calgary, and MICCAI challenges, with only around 7%, 5%, and 3.5% usage rates, respectively (see Figure 9c).

In addition, most studies reviewed by this study tested their proposed model using acceleration factor (R) $2 \leq R < 6$. The least simulated acceleration factor was $R \geq 12$ and $6 \leq R < 8$ with 13.4% and 7.3% of usage, respectively. The percentage of acceleration usage is summarized in Figure 9d.

## 4.2 Clinical evaluations

The metrics presented in Section 4.1 quantify the quality of image reconstruction, but their results may not directly correlate with clinical outcomes. Several studies have been conducted to evaluate the clinical significance of CS-MRI using DL models. For example, a study found that DL-based CS-MRI and fully sampled MRI images showed no significant differences (p-values > 0.05) in the organ-based image quality of the liver, pancreas, spleen, and kidneys, number and measured diameter of the detected lesions while reducing the imaging time by more than 85% (Herrmann et al., 2023).

Similarly, another study showed that brain MRI images accelerated up to $4\times$ and $14\times$ had sufficient image quality for diagnostic and screening purposes, respectively (Radmanesh et al., 2022). A third study found that there was no statistical significance ($p$-value = 0.521) between the DL-based T2-FLAIR MRI image and standard T1c MRI images in the assessment of inflammatory knee synovitis (Feuerriegel et al., 2023). These studies are consistent with another conducted to compare the diagnostic performance of DL MRI and standard MRI images in detecting knee abnormalities (Johnson et al., 2023).

A recent study found that there is no significant difference in the overall quality of MRI images generated by DL and standard fully sampled images for various MRI sequences, including T2 and diffusion-weighted imaging, for patients with prostate cancer. The study also revealed that DL MRI and standard MR images identified a similar number of Prostate Imaging Reporting and Data System (PIRADS) $\geq 3$ lesions. However, the imaging time was significantly reduced by about 3.7-fold with the use of DL MRI. This study's findings suggest that DL MRI can be a viable alternative to standard MRI imaging for prostate cancer patients, as it can produce similar quality images in a significantly shorter acquisition time (Johnson et al., 2022).

# 5 Discussion

The rapid advancement of DL in the field of computer vision has led to a significant increase in the number of studies utilizing DL to solve CS-MRI, as depicted in Figure 7. In this study, we provide a comprehensive overview of DL models and training approaches, including the use of GAN and neural architecture such as U-net and vision transformers for an end-to-end approach. Recent studies have extensively used the unroll optimization and DC layers due to their advantages, allowing for the employment of smaller ConvNets. This, in turn, reduces the number of trainable parameters, minimizing the risk of overfitting. Moreover, our study also revealed that the majority of studies employ SSIM, PSNR, and NMSE metrics to measure the image quality of predicted images.

The objective of DL-based CS-MRI approaches is to decrease imaging time, thereby enhancing throughput and minimizing the chances of patient movement. Apart from this, other methods, such as super-resolution and image synthesis, can also be adopted to decrease imaging time. These approaches strive to enhance the resolution of images obtained from lower spatial resolution (Chang et al., 2024; Xie et al., 2022) and produce MRI images of high spatial resolution and signal-to-noise ratio from lower $B_0$ (Eidex et al., 2023). These techniques can significantly improve MRI imaging with permanent magnets having low $B_0$, ultimately increasing the image quality. Portable MRI scanners widely use permanent magnets, which substantially reduce maintenance costs and make them more suitable for low-income and middle-income countries. Although these techniques can reduce the need for stronger gradient magnets and minimize patients' nerve stimulation, they were not included in this systematic review since they did not employ compressed sensing algorithms to train their models.

The coil sensitivity map, depicted in Figure 2b, is essential to consider the non-uniform sensitivity of the receiver coils. A precise sensitivity map is crucial for generating consistent and accurate MRI images and quantitative MRI maps across different hospitals. Most of the reviewed studies use the ESPIRiT algorithm (Uecker et al., 2014) provided by the Berkeley

Advanced Reconstruction Toolbox (https://mrirecon.github.io/bart/) to pre-calculate the sensitivity map. However, this algorithm involves significant computation that may hinder its application in MRI-guided surgery and treatment, where rapid image reconstruction is required. We anticipate that DL-based CS-MRI techniques, which can simultaneously predict coil sensitivity maps and coil images (Feng et al., 2023; Sun et al., 2023; Z. Wang et al., 2024), will be explored further in the future, particularly for MRI-guided treatment methods such as MRI-guided adaptive radiation therapy. Furthermore, most CS-MRI methods require prior knowledge of the sampling pattern, which may not be available to the user in real clinical applications. Therefore, an approach that can handle deviations from the actual sampling mask or jointly predict the optimal sampling pattern and predict the images (Seo et al., 2022) is required.

In a recent systematic review by Hu, Mingzhe, *et al.*, the use of language models in medical imaging was explored in detail (Hu et al., 2024). However, our comprehensive review has found that the utilization of foundational models in training DL-based CS-MRI is currently quite limited. We believe that these models could have a more significant impact on DL-based CS-MRI. By providing prior knowledge about the sampling pattern without explicitly specifying the sampling images, these models could be extremely helpful. Additionally, their prior inputs about the MRI sequence and imaging region could aid in training a model specific to that region and sequence, potentially improving the reconstruction of out-of-distribution and out-of-region images.

## 6 Concluding remarks

The current DL-based CS-MRI models are usually trained by using 2D fastMRI datasets, limiting the spatial resolution for small lesion detection. With the advancement of MRI techniques, future research can significantly benefit from the availability of large 3D or 4D raw *k*-space datasets featuring abnormalities. Such datasets can enable the development of tailored 3D models, potentially allowing real-time tumor tracking during radiation therapy for patients with conditions like lung cancer. By incorporating high-dimensional datasets, it would be possible to report accurate clinical endpoints to enhance cancer prognosis.

## Acknowledgments

This research is supported in part by the National Institutes of Health under Award Number R56EB033332, R01EB032680, and R01CA272991.

## Conflicts of interest

There are no conflicts of interest declared by the authors.

## References

Aggarwal, H.K., Mani, M.P., Jacob, M., 2018. MoDL: Model-based deep learning architecture for inverse problems. IEEE transactions on medical imaging 38, 394–405.

Avidan, N., Freiman, M., 2024. MA-RECON: Mask-aware deep-neural-network for robust fast MRI k-space interpolation. Computer Methods and Programs in Biomedicine 244, 107942.

Ayde, R., Senft, T., Salameh, N., Sarracanie, M., 2022. Deep learning for fast low-field MRI acquisitions. Scientific reports 12, 11394.

Balakrishnan, G., Zhao, A., Sabuncu, M.R., Guttag, J., Dalca, A.V., 2019. Voxelmorph: a learning framework for deformable medical image registration. IEEE transactions on medical imaging 38, 1788–1800.

Beauferris, Y., Teuwen, J., Karkalousos, D., Moriakov, N., Caan, M., Yiasemis, G., Rodrigues, L., Lopes, A., Pedrini, H., Rittner, L., others, 2022. Multi-coil mri reconstruction challenge—assessing brain mri reconstruction models and their generalizability to varying coil configurations. Frontiers in Neuroscience 16, 919186.

Bustin, A., Fuin, N., Botnar, R.M., Prieto, C., 2020. From Compressed-Sensing to Artificial Intelligence-Based Cardiac MRI Reconstruction. Front. Cardiovasc. Med. 7, 17. https://doi.org/10.3389/fcvm.2020.00017

Cao, C., Cui, Z.-X., Wang, Y., Liu, S., Chen, T., Zheng, H., Liang, D., Zhu, Y., 2024. High-Frequency Space Diffusion Model for Accelerated MRI. IEEE Transactions on Medical Imaging.

Cao, J., Liu, S., Liu, H., Lu, H., 2020. CS-MRI reconstruction based on analysis dictionary learning and manifold structure regularization. Neural Networks 123, 217–233.

Chan, S.H., 2024. Tutorial on Diffusion Models for Imaging and Vision. arXiv preprint arXiv:2403.18103.

Chang, C.-W., Peng, J., Safari, M., Salari, E., Pan, S., Roper, J., Qiu, R.L., Gao, Y., Shu, H.-K., Mao, H., others, 2024. High-resolution MRI synthesis using a data-driven framework with denoising diffusion probabilistic modeling. Physics in Medicine & Biology 69, 045001.

Chen, E.Z., Wang, P., Chen, X., Chen, T., Sun, S., 2022. Pyramid convolutional RNN for MRI image reconstruction. IEEE Transactions on Medical Imaging 41, 2033–2047.

Chen, Y., Schonlieb, C.-B., Lio, P., Leiner, T., Dragotti, P.L., Wang, G., Rueckert, D., Firmin, D., Yang, G., 2022. AI-Based Reconstruction for Fast MRI—A Systematic Review and Meta-Analysis. Proc. IEEE 110, 224–245. https://doi.org/10.1109/JPROC.2022.3141367

Cheng, J., Cui, Z.-X., Huang, W., Ke, Z., Ying, L., Wang, H., Zhu, Y., Liang, D., 2021. Learning data consistency and its application to dynamic MR imaging. IEEE Transactions on Medical Imaging 40, 3140–3153.

Chung, H., Ye, J.C., 2022. Score-based diffusion models for accelerated MRI. Medical image analysis 80, 102479.

Dai, Y., Wang, C., Wang, H., 2023. Deep compressed sensing MRI via a gradient-enhanced fusion model. Medical Physics 50, 1390–1405.

Dawood, P., Breuer, F., Stebani, J., Burd, P., Homolya, I., Oberberger, J., Jakob, P.M., Blaimer, M., 2023. Iterative training of robust k-space interpolation networks for improved image reconstruction with limited scan specific training samples. Magnetic Resonance in Medicine 89, 812–827.
Desai, A.D., Ozturkler, B.M., Sandino, C.M., Boutin, R., Willis, M., Vasanawala, S., Hargreaves, B.A., Ré, C., Pauly, J.M., Chaudhari, A.S., 2023. Noise2Recon: Enabling SNR-robust MRI reconstruction with semi-supervised and self-supervised learning. Magnetic Resonance in Medicine 90, 2052–2070.
Deshmane, A., Gulani, V., Griswold, M.A., Seiberlich, N., 2012. Parallel MR imaging. Magnetic Resonance Imaging 36, 55–72. https://doi.org/10.1002/jmri.23639
Do, W.-J., Seo, S., Han, Y., Ye, J.C., Choi, S.H., Park, S.-H., 2020. Reconstruction of multicontrast MR images through deep learning. Medical physics 47, 983–997.
Dong, X., Lei, Y., Wang, T., Thomas, M., Tang, L., Curran, W.J., Liu, T., Yang, X., 2019. Automatic multiorgan segmentation in thorax CT images using U-net-GAN. Medical physics 46, 2157–2168.
Donoho, D.L., 2006. For most large underdetermined systems of linear equations, the minimal $\ell 1$ solution is also the sparsest solution. Communications on pure and applied mathematics 59, 797–829.
Dosovitskiy, A., Beyer, L., Kolesnikov, A., Weissenborn, D., Zhai, X., Unterthiner, T., Dehghani, M., Minderer, M., Heigold, G., Gelly, S., others, 2020. An image is worth 16x16 words: Transformers for image recognition at scale. arXiv preprint arXiv:2010.11929.
Duan, C., Xiong, Y., Cheng, K., Xiao, S., Lyu, J., Wang, C., Bian, X., Zhang, J., Zhang, D., Chen, L., others, 2022. Accelerating susceptibility-weighted imaging with deep learning by complex-valued convolutional neural network (ComplexNet): validation in clinical brain imaging. European Radiology 32, 5679–5687.
Eidex, Z., Wang, J., Safari, M., Elder, E., Wynne, J., Wang, T., Shu, H.-K., Mao, H., Yang, X., 2023. High-resolution 3T to 7T MRI Synthesis with a Hybrid CNN-Transformer Model. arXiv preprint arXiv:2311.15044.
Ekanayake, M., Pawar, K., Harandi, M., Egan, G., Chen, Z., 2024. McSTRA: A multi-branch cascaded swin transformer for point spread function-guided robust MRI reconstruction. Computers in Biology and Medicine 168, 107775.
Elmas, G., Dar, S.U., Korkmaz, Y., Ceyani, E., Susam, B., Ozbey, M., Avestimehr, S., Çukur, T., 2022. Federated learning of generative image priors for MRI reconstruction. IEEE Transactions on Medical Imaging.
Fang, J., Lin, H., Chen, X., Zeng, K., 2022. A hybrid network of cnn and transformer for lightweight image super-resolution, in: Proceedings of the IEEE/CVF Conference on Computer Vision and Pattern Recognition. pp. 1103–1112.
Feng, C.-M., Fu, H., Yuan, S., Xu, Y., 2021. Multi-contrast MRI super-resolution via a multi-stage integration network, in: Medical Image Computing and Computer Assisted Intervention–MICCAI 2021: 24th International Conference, Strasbourg, France, September 27–October 1, 2021, Proceedings, Part VI 24. Springer, pp. 140–149.
Feng, C.-M., Yan, Y., Wang, S., Xu, Y., Shao, L., Fu, H., 2022. Specificity-preserving federated learning for MR image reconstruction. IEEE Transactions on Medical Imaging.
Feng, R., Wu, Q., Feng, J., She, H., Liu, C., Zhang, Y., Wei, H., 2023. IMJENSE: scan-specific implicit representation for joint coil sensitivity and image estimation in parallel MRI. IEEE Transactions on Medical Imaging.

Feuerriegel, G.C., Goller, S.S., von Deuster, C., Sutter, R., 2023. Inflammatory Knee Synovitis: Evaluation of an Accelerated FLAIR Sequence Compared With Standard Contrast-Enhanced Imaging. Investigative Radiology 10–1097.

Gao, Z., Guo, Y., Zhang, J., Zeng, T., Yang, G., 2023. Hierarchical perception adversarial learning framework for compressed sensing MRI. IEEE Transactions on Medical Imaging.

Geng, C., Jiang, M., Fang, X., Li, Y., Jin, G., Chen, A., Liu, F., 2023. HFIST-Net: High-throughput fast iterative shrinkage thresholding network for accelerating MR image reconstruction. Computer Methods and Programs in Biomedicine 232, 107440.

Goodfellow, I., Pouget-Abadie, J., Mirza, M., Xu, B., Warde-Farley, D., Ozair, S., Courville, A., Bengio, Y., 2020. Generative adversarial networks. Communications of the ACM 63, 139–144.

Güngör, A., Dar, S.U., Öztürk, Ş., Korkmaz, Y., Bedel, H.A., Elmas, G., Ozbey, M., Çukur, T., 2023. Adaptive diffusion priors for accelerated MRI reconstruction. Medical Image Analysis 88, 102872.

Guo, D., Zeng, G., Fu, H., Wang, Z., Yang, Y., Qu, X., 2023. A Joint Group Sparsity-based deep learning for multi-contrast MRI reconstruction. Journal of Magnetic Resonance 346, 107354.

Guo, P., Mei, Y., Zhou, J., Jiang, S., Patel, V.M., 2023. ReconFormer: Accelerated MRI reconstruction using recurrent transformer. IEEE transactions on medical imaging.

Guo, P., Valanarasu, J.M.J., Wang, P., Zhou, J., Jiang, S., Patel, V.M., 2021a. Over-and-under complete convolutional rnn for mri reconstruction, in: Medical Image Computing and Computer Assisted Intervention–MICCAI 2021: 24th International Conference, Strasbourg, France, September 27–October 1, 2021, Proceedings, Part VI 24. Springer, pp. 13–23.

Guo, P., Wang, P., Zhou, J., Jiang, S., Patel, V.M., 2021b. Multi-Institutional Collaborations for Improving Deep Learning-Based Magnetic Resonance Image Reconstruction Using Federated Learning, in: Proceedings of the IEEE/CVF Conference on Computer Vision and Pattern Recognition (CVPR). pp. 2423–2432.

Haji-Valizadeh, H., Rahsepar, A.A., Collins, J.D., Bassett, E., Isakova, T., Block, T., Adluru, G., DiBella, E.V., Lee, D.C., Carr, J.C., others, 2018. Validation of highly accelerated real-time cardiac cine MRI with radial k-space sampling and compressed sensing in patients at 1.5 T and 3T. Magnetic resonance in medicine 79, 2745–2751.

Haldar, J.P., Hernando, D., Liang, Z.-P., 2010. Compressed-sensing MRI with random encoding. IEEE transactions on Medical Imaging 30, 893–903.

Hammernik, K., Klatzer, T., Kobler, E., Recht, M.P., Sodickson, D.K., Pock, T., Knoll, F., 2018. Learning a variational network for reconstruction of accelerated MRI data. Magnetic resonance in medicine 79, 3055–3071.

Han, X., 2017. MR-based synthetic CT generation using a deep convolutional neural network method. Medical physics 44, 1408–1419.

He, K., Zhang, X., Ren, S., Sun, J., 2016. Deep residual learning for image recognition, in: Proceedings of the IEEE Conference on Computer Vision and Pattern Recognition. pp. 770–778.

Herrmann, J., Wessling, D., Nickel, D., Arberet, S., Almansour, H., Afat, C., Afat, S., Gassenmaier, S., Othman, A.E., 2023. Comprehensive clinical evaluation of a deep learning-accelerated, single-breath-hold abdominal HASTE at 1.5 T and 3 T. Academic Radiology 30, 93–102.
Heydari, A., Ahmadi, A., Kim, T.H., Bilgic, B., 2024. Joint MAPLE: Accelerated joint T1 and T 2* T _2^ * mapping with scan-specific self-supervised networks. Magnetic Resonance in Medicine.
Hong, G.Q., Wei, Y.T., Morley, W.A., Wan, M., Mertens, A.J., Su, Y., Cheng, H.-L.M., 2023. Dual-domain accelerated MRI reconstruction using transformers with learning-based undersampling. Computerized Medical Imaging and Graphics 106, 102206.
Hricak, H., Abdel-Wahab, M., Atun, R., Lette, M.M., Paez, D., Brink, J.A., Donoso-Bach, L., Frija, G., Hierath, M., Holmberg, O., others, 2021. Medical imaging and nuclear medicine: a Lancet Oncology Commission. The Lancet Oncology 22, e136–e172.
Hu, C., Li, C., Wang, H., Liu, Q., Zheng, H., Wang, S., 2021. Self-supervised learning for mri reconstruction with a parallel network training framework, in: Medical Image Computing and Computer Assisted Intervention–MICCAI 2021: 24th International Conference, Strasbourg, France, September 27–October 1, 2021, Proceedings, Part VI 24. Springer, pp. 382–391.
Hu, M., Qian, J.Y., Pan, S., Li, Y., Qiu, R.L., Yang, X., 2024. Advancing medical imaging with language models: featuring a spotlight on ChatGPT. Physics in Medicine and Biology.
Hu, Y., Xu, Y., Tian, Q., Chen, F., Shi, X., Moran, C.J., Daniel, B.L., Hargreaves, B.A., 2021. RUN-UP: Accelerated multishot diffusion-weighted MRI reconstruction using an unrolled network with U-Net as priors. Magnetic resonance in medicine 85, 709–720.
Huang, G., Liu, Z., Van Der Maaten, L., Weinberger, K.Q., 2017. Densely connected convolutional networks, in: Proceedings of the IEEE Conference on Computer Vision and Pattern Recognition. pp. 4700–4708.
Huang, J., Fang, Y., Wu, Y., Wu, H., Gao, Z., Li, Y., Del Ser, J., Xia, J., Yang, G., 2022. Swin transformer for fast MRI. Neurocomputing 493, 281–304.
Hyun, C.M., Kim, H.P., Lee, S.M., Lee, S., Seo, J.K., 2018. Deep learning for undersampled MRI reconstruction. Physics in Medicine & Biology 63, 135007.
Išgum, I., Benders, M.J., Avants, B., Cardoso, M.J., Counsell, S.J., Gomez, E.F., Gui, L., H\Huppi, P.S., Kersbergen, K.J., Makropoulos, A., others, 2015. Evaluation of automatic neonatal brain segmentation algorithms: the NeoBrainS12 challenge. Medical image analysis 20, 135–151.
Isola, P., Zhu, J.-Y., Zhou, T., Efros, A.A., 2017. Image-to-image translation with conditional adversarial networks, in: Proceedings of the IEEE Conference on Computer Vision and Pattern Recognition. pp. 1125–1134.
Jiang, S., Eberhart, C.G., Lim, M., Heo, H.-Y., Zhang, Y., Blair, L., Wen, Z., Holdhoff, M., Lin, D., Huang, P., others, 2019. Identifying recurrent malignant glioma after treatment using amide proton transfer-weighted MR imaging: a validation study with image-guided stereotactic biopsy. Clinical Cancer Research 25, 552–561.
Jin, Z., Xiang, Q.-S., 2023. Improving accelerated MRI by deep learning with sparsified complex data. Magnetic Resonance in Medicine 89, 1825–1838.
Johnson, P.M., Lin, D.J., Zbontar, J., Zitnick, C.L., Sriram, A., Muckley, M., Babb, J.S., Kline, M., Ciavarra, G., Alaia, E., others, 2023. Deep learning reconstruction enables prospectively accelerated clinical knee MRI. Radiology 307, e220425.
Johnson, P.M., Tong, A., Donthireddy, A., Melamud, K., Petrocelli, R., Smereka, P., Qian, K., Keerthivasan, M.B., Chandarana, H., Knoll, F., 2022. Deep learning reconstruction enables highly accelerated biparametric MR imaging of the prostate. Journal of Magnetic Resonance Imaging 56, 184–195.
Karras, T., Laine, S., Aila, T., 2019. A style-based generator architecture for generative adversarial networks, in: Proceedings of the IEEE/CVF Conference on Computer Vision and Pattern Recognition. pp. 4401–4410.
Khawaled, S., Freiman, M., 2024. NPB-REC: A non-parametric Bayesian deep-learning approach for undersampled MRI reconstruction with uncertainty estimation. Artificial Intelligence in Medicine 102798.
Kim, M., Chung, W., 2022. A cascade of preconditioned conjugate gradient networks for accelerated magnetic resonance imaging. Computer Methods and Programs in Biomedicine 225, 107090.
Kofler, A., Pali, M.-C., Schaeffter, T., Kolbitsch, C., 2023. Deep supervised dictionary learning by algorithm unrolling—Application to fast 2D dynamic MR image reconstruction. Medical Physics 50, 2939–2960.
Korkmaz, Y., Cukur, T., Patel, V.M., 2023. Self-supervised MRI reconstruction with unrolled diffusion models, in: International Conference on Medical Image Computing and Computer-Assisted Intervention. Springer, pp. 491–501.
Krizhevsky, A., Sutskever, I., Hinton, G.E., 2012. Imagenet classification with deep convolutional neural networks. Advances in neural information processing systems 25.
Larkman, D.J., Nunes, R.G., 2007. Parallel magnetic resonance imaging. Phys. Med. Biol. 52, R15–R55. https://doi.org/10.1088/0031-9155/52/7/R01
Lauterbur, P.C., 1973. Image Formation by Induced Local Interactions: Examples Employing Nuclear Magnetic Resonance. Nature 242, 190–191. https://doi.org/10.1038/242190a0
LeCun, Y., Bengio, Y., others, 1995. Convolutional networks for images, speech, and time series. The handbook of brain theory and neural networks 3361, 1995.
Levac, B., Jalal, A., Tamir, J.I., 2023. Accelerated motion correction for MRI using score-based generative models, in: 2023 IEEE 20th International Symposium on Biomedical Imaging (ISBI). IEEE, pp. 1–5.
Levac, B.R., Arvinte, M., Tamir, J.I., 2023. Federated end-to-end unrolled models for magnetic resonance image reconstruction. Bioengineering 10, 364.
Li, H., Xu, Z., Taylor, G., Studer, C., Goldstein, T., 2018. Visualizing the loss landscape of neural nets. Advances in neural information processing systems 31.
Li, H., Yang, M., Kim, J.H., Zhang, C., Liu, R., Huang, P., Liang, D., Zhang, X., Li, X., Ying, L., 2023. SuperMAP: Deep ultrafast MR relaxometry with joint spatiotemporal undersampling. Magnetic resonance in medicine 89, 64–76.
Li, T., Sahu, A.K., Zaheer, M., Sanjabi, M., Talwalkar, A., Smith, V., 2020. Federated optimization in heterogeneous networks. Proceedings of Machine learning and systems 2, 429–450.
Li, Y., Joaquim, M.R., Pickup, S., Song, H.K., Zhou, R., Fan, Y., 2024. Learning ADC maps from accelerated radial k-space diffusion-weighted MRI in mice using a deep CNN-transformer model. Magnetic Resonance in Medicine 91, 105–117.
Liu, F., Kijowski, R., El Fakhri, G., Feng, L., 2021. Magnetic resonance parameter mapping using model-guided self-supervised deep learning. Magnetic resonance in medicine 85, 3211–3226.

Liu, Q., Yang, Q., Cheng, H., Wang, S., Zhang, M., Liang, D., 2020. Highly undersampled magnetic resonance imaging reconstruction using autoencoding priors. Magnetic resonance in medicine 83, 322–336.
Liu, R., Zhang, Y., Cheng, S., Luo, Z., Fan, X., 2020. A deep framework assembling principled modules for CS-MRI: unrolling perspective, convergence behaviors, and practical modeling. IEEE Transactions on Medical Imaging 39, 4150–4163.
Liu, S., Li, H., Liu, Y., Cheng, G., Yang, G., Wang, H., Zheng, H., Liang, D., Zhu, Y., 2022. Highly accelerated MR parametric mapping by undersampling the k-space and reducing the contrast number simultaneously with deep learning. Physics in Medicine & Biology 67, 185004.
Liu, S., Liu, Y., Xu, X., Chen, R., Liang, D., Jin, Q., Liu, H., Chen, G., Zhu, Y., 2023. Accelerated cardiac diffusion tensor imaging using deep neural network. Physics in Medicine & Biology 68, 025008.
Liu, X., Wang, J., Sun, H., Chandra, S.S., Crozier, S., Liu, F., 2021. On the regularization of feature fusion and mapping for fast MR multi-contrast imaging via iterative networks. Magnetic resonance imaging 77, 159–168.
Liu, Y., Leong, A.T., Zhao, Y., Xiao, L., Mak, H.K., Tsang, A.C.O., Lau, G.K., Leung, G.K., Wu, E.X., 2021. A low-cost and shielding-free ultra-low-field brain MRI scanner. Nature communications 12, 7238.
Liu, Z., Lin, Y., Cao, Y., Hu, H., Wei, Y., Zhang, Z., Lin, S., Guo, B., 2021. Swin transformer: Hierarchical vision transformer using shifted windows, in: Proceedings of the IEEE/CVF International Conference on Computer Vision. pp. 10012–10022.
Lu, T., Zhang, X., Huang, Y., Guo, D., Huang, F., Xu, Q., Hu, Y., Ou-Yang, L., Lin, J., Yan, Z., others, 2020. pFISTA-SENSE-ResNet for parallel MRI reconstruction. Journal of Magnetic Resonance 318, 106790.
Lyu, J., Tian, Y., Cai, Q., Wang, C., Qin, J., 2023. Adaptive channel-modulated personalized federated learning for magnetic resonance image reconstruction. Computers in Biology and Medicine 165, 107330.
Lyu, Q., Shan, H., Steber, C., Helis, C., Whitlow, C., Chan, M., Wang, G., 2020. Multi-contrast super-resolution MRI through a progressive network. IEEE transactions on medical imaging 39, 2738–2749.
Maclaren, J., Herbst, M., Speck, O., Zaitsev, M., 2013. Prospective motion correction in brain imaging: a review. Magnetic resonance in medicine 69, 621–636.
Mardani, M., Gong, E., Cheng, J.Y., Vasanawala, S.S., Zaharchuk, G., Xing, L., Pauly, J.M., 2018. Deep generative adversarial neural networks for compressive sensing MRI. IEEE transactions on medical imaging 38, 167–179.
McMahan, B., Moore, E., Ramage, D., Hampson, S., y Arcas, B.A., 2017. Communication-efficient learning of deep networks from decentralized data, in: Artificial Intelligence and Statistics. PMLR, pp. 1273–1282.
Mendrik, A.M., Vincken, K.L., Kuijf, H.J., Breeuwer, M., Bouvy, W.H., De Bresser, J., Alansary, A., De Bruijne, M., Carass, A., El-Baz, A., others, 2015. MRBrainS challenge: online evaluation framework for brain image segmentation in 3T MRI scans. Computational intelligence and neuroscience 2015, 1–1.
Meng, Z., Guo, R., Li, Yudu, Guan, Y., Wang, T., Zhao, Y., Sutton, B., Li, Yao, Liang, Z.-P., 2021. Accelerating T2 mapping of the brain by integrating deep learning priors with low-rank and sparse modeling. Magnetic Resonance in Medicine 85, 1455–1467.

Menze, B.H., Jakab, A., Bauer, S., Kalpathy-Cramer, J., Farahani, K., Kirby, J., Burren, Y., Porz, N., Slotboom, J., Wiest, R., others, 2014. The multimodal brain tumor image segmentation benchmark (BRATS). IEEE transactions on medical imaging 34, 1993–2024.
Mirza, M., Osindero, S., 2014. Conditional generative adversarial nets. arXiv preprint arXiv:1411.1784.
Murali, S., Ding, H., Adedeji, F., Qin, C., Obungoloch, J., Asllani, I., Anazodo, U., Ntusi, N.A., Mammen, R., Niendorf, T., others, 2023. Bringing MRI to low-and middle-income countries: directions, challenges and potential solutions. NMR in Biomedicine e4992.
Murugesan, B., Ramanarayanan, S., Vijayarangan, S., Ram, K., Jagannathan, N.R., Sivaprakasam, M., 2021. A deep cascade of ensemble of dual domain networks with gradient-based T1 assistance and perceptual refinement for fast MRI reconstruction. Computerized Medical Imaging and Graphics 91, 101942.
Narnhofer, D., Hammernik, K., Knoll, F., Pock, T., 2019. Inverse GANs for accelerated MRI reconstruction, in: Wavelets and Sparsity XVIII. SPIE, pp. 381–392.
Pan, S., Abouei, E., Wynne, J., Chang, C.-W., Wang, T., Qiu, R.L., Li, Y., Peng, J., Roper, J., Patel, P., others, 2023a. Synthetic CT generation from MRI using 3D transformer-based denoising diffusion model. Medical Physics.
Pan, S., Wang, T., Qiu, R.L., Axente, M., Chang, C.-W., Peng, J., Patel, A.B., Shelton, J., Patel, S.A., Roper, J., others, 2023b. 2D medical image synthesis using transformer-based denoising diffusion probabilistic model. Physics in Medicine & Biology 68, 105004.
Pramanik, A., Bhave, S., Sajib, S., Sharma, S.D., Jacob, M., 2023. Adapting model-based deep learning to multiple acquisition conditions: Ada-MoDL. Magnetic resonance in medicine 90, 2033–2051.
Qiao, X., Huang, Y., Li, W., 2023. MEDL-Net: A model-based neural network for MRI reconstruction with enhanced deep learned regularizers. Magnetic Resonance in Medicine 89, 2062–2075.
Qin, C., Duan, J., Hammernik, K., Schlemper, J., Küstner, T., Botnar, R., Prieto, C., Price, A.N., Hajnal, J.V., Rueckert, D., 2021. Complementary time-frequency domain networks for dynamic parallel MR image reconstruction. Magnetic Resonance in Medicine 86, 3274–3291.
Qin, C., Schlemper, J., Caballero, J., Price, A.N., Hajnal, J.V., Rueckert, D., 2018. Convolutional recurrent neural networks for dynamic MR image reconstruction. IEEE transactions on medical imaging 38, 280–290.
Qu, B., Zhang, J., Kang, T., Lin, J., Lin, M., She, H., Wu, Q., Wang, M., Zheng, G., 2024. Radial magnetic resonance image reconstruction with a deep unrolled projected fast iterative soft-thresholding network. Computers in Biology and Medicine 168, 107707.
Quan, T.M., Nguyen-Duc, T., Jeong, W.-K., 2018. Compressed sensing MRI reconstruction using a generative adversarial network with a cyclic loss. IEEE transactions on medical imaging 37, 1488–1497.
Radmanesh, A., Muckley, M.J., Murrell, T., Lindsey, E., Sriram, A., Knoll, F., Sodickson, D.K., Lui, Y.W., 2022. Exploring the acceleration limits of deep learning variational network–based two-dimensional brain MRI. Radiology: Artificial Intelligence 4, e210313.

Ravishankar, S., Bresler, Y., 2010. MR image reconstruction from highly undersampled k-space data by dictionary learning. IEEE transactions on medical imaging 30, 1028–1041.
Rohlfing, T., Zahr, N.M., Sullivan, E.V., Pfefferbaum, A., 2010. The SRI24 multichannel atlas of normal adult human brain structure. Human brain mapping 31, 798–819.
Ronneberger, O., Fischer, P., Brox, T., 2015. U-net: Convolutional networks for biomedical image segmentation, in: Medical Image Computing and Computer-Assisted Intervention–MICCAI 2015: 18th International Conference, Munich, Germany, October 5-9, 2015, Proceedings, Part III 18. Springer, pp. 234–241.
Ryu, K., Lee, J.-H., Nam, Y., Gho, S.-M., Kim, H.-S., Kim, D.-H., 2021. Accelerated multicontrast reconstruction for synthetic MRI using joint parallel imaging and variable splitting networks. Medical physics 48, 2939–2950.
Safari, M., Fatemi, A., Archambault, L., 2023a. MedFusionGAN: multimodal medical image fusion using an unsupervised deep generative adversarial network. BMC Medical Imaging 23, 203.
Safari, M., Yang, X., Fatemi, A., 2024. MRI data consistency guided conditional diffusion probabilistic model for MR imaging acceleration, in: Medical Imaging 2024: Clinical and Biomedical Imaging. SPIE, pp. 202–205.
Safari, M., Yang, X., Fatemi, A., Archambault, L., 2023b. MRI motion artifact reduction using a conditional diffusion probabilistic model (MAR-CDPM). Medical Physics.
Sandino, C.M., Lai, P., Vasanawala, S.S., Cheng, J.Y., 2021. Accelerating cardiac cine MRI using a deep learning-based ESPIRiT reconstruction. Magnetic Resonance in Medicine 85, 152–167.
Schlemper, J., Caballero, J., Hajnal, J.V., Price, A., Rueckert, D., 2017. A deep cascade of convolutional neural networks for MR image reconstruction, in: Information Processing in Medical Imaging: 25th International Conference, IPMI 2017, Boone, NC, USA, June 25-30, 2017, Proceedings 25. Springer, pp. 647–658.
Seo, S., Luu, H.M., Choi, S.H., Park, S.-H., 2022. Simultaneously optimizing sampling pattern for joint acceleration of multi-contrast MRI using model-based deep learning. Medical Physics 49, 5964–5980.
Shaul, R., David, I., Shitrit, O., Raviv, T.R., 2020. Subsampled brain MRI reconstruction by generative adversarial neural networks. Medical Image Analysis 65, 101747.
Shitrit, O., Riklin Raviv, T., 2017. Accelerated magnetic resonance imaging by adversarial neural network, in: Deep Learning in Medical Image Analysis and Multimodal Learning for Clinical Decision Support: Third International Workshop, DLMIA 2017, and 7th International Workshop, ML-CDS 2017, Held in Conjunction with MICCAI 2017, Québec City, QC, Canada, September 14, Proceedings 3. Springer, pp. 30–38.
Slavkova, K.P., DiCarlo, J.C., Wadhwa, V., Kumar, S., Wu, C., Virostko, J., Yankeelov, T.E., Tamir, J.I., 2023. An untrained deep learning method for reconstructing dynamic MR images from accelerated model-based data. Magnetic resonance in medicine 89, 1617–1633.
Slipsager, J.M., Glimberg, S.L., Søgaard, J., Paulsen, R.R., Johannesen, H.H., Martens, P.C., Seth, A., Marner, L., Henriksen, O.M., Olesen, O.V., others, 2020. Quantifying the financial savings of motion correction in brain MRI: a model-based estimate of the costs arising from patient head motion and potential savings from implementation of motion correction. Journal of Magnetic Resonance Imaging 52, 731–738.

Sohl-Dickstein, J., Weiss, E., Maheswaranathan, N., Ganguli, S., 2015. Deep unsupervised learning using nonequilibrium thermodynamics, in: International Conference on Machine Learning. PMLR, pp. 2256–2265.

Sun, K., Wang, Q., Shen, D., 2023. Joint Cross-Attention Network with Deep Modality Prior for Fast MRI Reconstruction. IEEE Transactions on Medical Imaging.

Sun, L., Fan, Z., Fu, X., Huang, Y., Ding, X., Paisley, J., 2019. A deep information sharing network for multi-contrast compressed sensing MRI reconstruction. IEEE Transactions on Image Processing 28, 6141–6153.

Sun, L., Huang, Y., Cai, C., Ding, X., 2017. Compressed sensing MRI using total variation regularization with K-space decomposition, in: 2017 IEEE International Conference on Image Processing (ICIP). IEEE, pp. 3061–3065.

Terpstra, M.L., Maspero, M., Verhoeff, J.J., van den Berg, C.A., 2023. Accelerated respiratory-resolved 4D-MRI with separable spatio-temporal neural networks. Medical physics 50, 5331–5342.

Uecker, M., Lai, P., Murphy, M.J., Virtue, P., Elad, M., Pauly, J.M., Vasanawala, S.S., Lustig, M., 2014. ESPIRiT—an eigenvalue approach to autocalibrating parallel MRI: where SENSE meets GRAPPA. Magnetic resonance in medicine 71, 990–1001.

Van Essen, D.C., Smith, S.M., Barch, D.M., Behrens, T.E., Yacoub, E., Ugurbil, K., Consortium, W.-M.H., others, 2013. The WU-Minn human connectome project: an overview. Neuroimage 80, 62–79.

Wang, B., Lian, Y., Xiong, X., Zhou, H., Liu, Z., Zhou, X., 2024. DCT-net: Dual-domain cross-fusion transformer network for MRI reconstruction. Magnetic Resonance Imaging 107, 69–79.

Wang, J., Awad, M., Zhou, R., Wang, Z., Wang, X., Feng, X., Yang, Y., Meyer, C., Kramer, C.M., Salerno, M., 2024. High-resolution spiral real-time cardiac cine imaging with deep learning-based rapid image reconstruction and quantification. NMR in Biomedicine 37, e5051.

Wang, K., Tamir, J.I., De Goyeneche, A., Wollner, U., Brada, R., Yu, S.X., Lustig, M., 2022. High fidelity deep learning-based MRI reconstruction with instance-wise discriminative feature matching loss. Magnetic Resonance in Medicine 88, 476–491.

Wang, S., Cheng, H., Ying, L., Xiao, T., Ke, Z., Zheng, H., Liang, D., 2020. DeepcomplexMRI: Exploiting deep residual network for fast parallel MR imaging with complex convolution. Magnetic resonance imaging 68, 136–147.

Wang, Z., Fang, H., Qian, C., Shi, B., Bao, L., Zhu, L., Zhou, J., Wei, W., Lin, J., Guo, D., others, 2024. A faithful deep sensitivity estimation for accelerated magnetic resonance imaging. IEEE Journal of Biomedical and Health Informatics.

Wang, Z., Qian, C., Guo, D., Sun, H., Li, R., Zhao, B., Qu, X., 2022. One-dimensional deep low-rank and sparse network for accelerated MRI. IEEE Transactions on Medical Imaging 42, 79–90.

Wang, Z., She, H., Zhang, Y., Du, Y.P., 2023. Parallel non-Cartesian spatial-temporal dictionary learning neural networks (stDLNN) for accelerating 4D-MRI. Medical image analysis 84, 102701.

Wen, J., Ahmad, R., Schniter, P., 2023. A conditional normalizing flow for accelerated multi-coil MR imaging. Proceedings of machine learning research 202, 36926.

Wu, Y., Ma, Y., Liu, J., Du, J., Xing, L., 2019. Self-attention convolutional neural network for improved MR image reconstruction. Information sciences 490, 317–328.

Wu, Z., Liao, W., Yan, C., Zhao, M., Liu, G., Ma, N., Li, X., 2023. Deep learning based MRI reconstruction with transformer. Computer Methods and Programs in Biomedicine 233, 107452.

Xiang, L., Chen, Y., Chang, W., Zhan, Y., Lin, W., Wang, Q., Shen, D., 2018. Ultra-fast t2-weighted mr reconstruction using complementary t1-weighted information, in: Medical Image Computing and Computer Assisted Intervention–MICCAI 2018: 21st International Conference, Granada, Spain, September 16-20, 2018, Proceedings, Part I. Springer, pp. 215–223.

Xie, H., Lei, Y., Wang, T., Roper, J., Dhabaan, A.H., Bradley, J.D., Liu, T., Mao, H., Yang, X., 2022. Synthesizing high-resolution magnetic resonance imaging using parallel cycle-consistent generative adversarial networks for fast magnetic resonance imaging. Medical Physics 49, 357–369.

Xie, Y., Li, Q., 2022. A Review of Deep Learning Methods for Compressed Sensing Image Reconstruction and Its Medical Applications. Electronics 11, 586. https://doi.org/10.3390/electronics11040586

Xu, J., Zu, T., Hsu, Y.-C., Wang, X., Chan, K.W., Zhang, Y., 2024. Accelerating CEST imaging using a model-based deep neural network with synthetic training data. Magnetic Resonance in Medicine 91, 583–599.

Xue, Y., Yu, J., Kang, H.S., Englander, S., Rosen, M.A., Song, H.K., 2012. Automatic coil selection for streak artifact reduction in radial MRI. Magnetic Resonance in Med 67, 470–476. https://doi.org/10.1002/mrm.23023

Yaman, B., Gu, H., Hosseini, S.A.H., Demirel, O.B., Moeller, S., Ellermann, J., Uğurbil, K., Akçakaya, M., 2022. Multi-mask self-supervised learning for physics-guided neural networks in highly accelerated magnetic resonance imaging. NMR in Biomedicine 35, e4798.

Yaman, B., Hosseini, S.A.H., Moeller, S., Ellermann, J., Uğurbil, K., Akçakaya, M., 2020. Self-supervised learning of physics-guided reconstruction neural networks without fully sampled reference data. Magnetic resonance in medicine 84, 3172–3191.

Yan, Y., Yang, T., Zhao, X., Jiao, C., Yang, A., Miao, J., 2023. DC-SiamNet: Deep contrastive Siamese network for self-supervised MRI reconstruction. Computers in Biology and Medicine 167, 107619.

Yang, G., Yu, S., Dong, H., Slabaugh, G., Dragotti, P.L., Ye, X., Liu, F., Arridge, S., Keegan, J., Guo, Y., others, 2017. DAGAN: deep de-aliasing generative adversarial networks for fast compressed sensing MRI reconstruction. IEEE transactions on medical imaging 37, 1310–1321.

Yang, Q., Liu, Y., Chen, T., Tong, Y., 2019. Federated machine learning: Concept and applications. ACM Transactions on Intelligent Systems and Technology (TIST) 10, 1–19.

Yang, Y., Sun, J., Li, H., Xu, Z., 2016. Deep ADMM-Net for compressive sensing MRI, in: Proceedings of the 30th International Conference on Neural Information Processing Systems. pp. 10–18.

Yiasemis, G., Sánchez, C.I., Sonke, J.-J., Teuwen, J., 2024. On retrospective k-space subsampling schemes for deep MRI reconstruction. Magnetic Resonance Imaging 107, 33–46.

Yoon, M.A., Gold, G.E., Chaudhari, A.S., 2023. Accelerated Musculoskeletal Magnetic Resonance Imaging. Magnetic Resonance Imaging jmri.29205. https://doi.org/10.1002/jmri.29205

Zbontar, J., Knoll, F., Sriram, A., Murrell, T., Huang, Z., Muckley, M.J., Defazio, A., Stern, R., Johnson, P., Bruno, M., others, 2018. fastMRI: An open dataset and benchmarks for accelerated MRI. arXiv preprint arXiv:1811.08839.

Zhao, X., Yang, T., Li, B., Zhang, X., 2023. SwinGAN: A dual-domain Swin Transformer-based generative adversarial network for MRI reconstruction. Computers in Biology and Medicine 153, 106513.

Zhou, B., Schlemper, J., Dey, N., Salehi, S.S.M., Sheth, K., Liu, C., Duncan, J.S., Sofka, M., 2022. Dual-domain self-supervised learning for accelerated non-Cartesian MRI reconstruction. Medical Image Analysis 81, 102538.

Zhou, Z., Han, F., Ghodrati, V., Gao, Y., Yin, W., Yang, Y., Hu, P., 2019. Parallel imaging and convolutional neural network combined fast MR image reconstruction: Applications in low-latency accelerated real-time imaging. Medical physics 46, 3399–3413.

Zhu, J.-Y., Park, T., Isola, P., Efros, A.A., 2017. Unpaired image-to-image translation using cycle-consistent adversarial networks, in: Proceedings of the IEEE International Conference on Computer Vision. pp. 2223–2232.

Zufiria, B., Qiu, S., Yan, K., Zhao, R., Wang, R., She, H., Zhang, C., Sun, B., Herman, P., Du, Y., others, 2022. A feature-based convolutional neural network for reconstruction of interventional MRI. NMR in Biomedicine 35, e4231.